\documentclass[aps,prx,twocolumn,superscriptaddress]{revtex4}
\usepackage{graphics}
\usepackage[dvips]{graphicx}
\usepackage{times}
\usepackage{multirow}
\usepackage{float}
\usepackage[T1]{fontenc}
\usepackage{color}
\usepackage[paperwidth=8.5in,paperheight=11in,centering,hmargin=0.79in,vmargin=0.79in]{geometry}

\usepackage{blkarray}
\usepackage{amsmath,bm}
\begin{document}

\title{A novel route to cyclic dominance in voluntary social dilemmas}

\author{Hao Guo}
\affiliation{School of Mechanical Engineering, Northwestern Polytechnical University, Xi'an, 710072 China}
\affiliation{Center for OPTical IMagery Analysis and Learning (OPTIMAL), Northwestern Polytechnical University, Xi'an, 710072 China}

\author{Zhao Song}
\affiliation{School of Mechanical Engineering, Northwestern Polytechnical University, Xi'an, 710072 China}
\affiliation{Center for OPTical IMagery Analysis and Learning (OPTIMAL), Northwestern Polytechnical University, Xi'an, 710072 China}

\author{Sun\v{c}ana Ge\v{c}ek}
\affiliation{Division for Marine and Environmental Research, Ru{\dj}er Bo\v{s}kovi\'{c} Institute, HR-10002 Zagreb, Croatia}

\author{Xuelong Li}
\affiliation{Center for OPTical IMagery Analysis and Learning (OPTIMAL), Northwestern Polytechnical University, Xi'an, 710072 China}
\affiliation{School of Computer Science, Northwestern Polytechnical University, Xi'an, 710072 China}

\author{Marko Jusup}
\thanks{Corresponding author: mjusup@gmail.com}
\affiliation{Tokyo Tech World Research Hub Initiative (WRHI), Institute of Innovative Research, Tokyo Institute of Technology, Tokyo, 152-8550 Japan}

\author{Matja{\v z} Perc}
\affiliation{Faculty of Natural Sciences and Mathematics, University of Maribor, Koro{\v s}ka cesta 160, 2000 Maribor, Slovenia}
\affiliation{Complexity Science Hub Vienna, Josefst\"{a}dterstra{\ss}e 39, 1080 Vienna, Austria}
\affiliation{Department of Medical Research, China Medical University Hospital, China Medical University, Taichung, Taiwan}

\author{Yamir Moreno}
\affiliation{Institute for Biocomputation and Physics of Complex Systems (BIFI), University of Zaragoza, Zaragoza 50009, Spain}
\affiliation{ISI Foundation, Turin 10126, Italy}

\author{Stefano Boccaletti}
\affiliation{Unmanned Systems Research Institute, Northwestern Polytechnical University, Xi'an, 710072 China}
\affiliation{CNR-Institute of Complex Systems, Via Madonna del Piano, 10, 50019 Sesto Fiorentino, Florence, Italy}
\affiliation{Moscow Institute of Physics and Technology, National Research University, Moscow Region, 141701 Russia}

\author{Zhen Wang}
\thanks{Corresponding author: zhenwang0@gmail.com}
\affiliation{School of Mechanical Engineering, Northwestern Polytechnical University, Xi'an, 710072 China}
\affiliation{Center for OPTical IMagery Analysis and Learning (OPTIMAL), Northwestern Polytechnical University, Xi'an, 710072 China}

\begin{abstract}
Cooperation is the backbone of modern human societies, making it a priority to understand how successful cooperation-sustaining mechanisms operate. Cyclic dominance, a non-transitive setup comprising at least three strategies wherein the first strategy overrules the second which overrules the third which, in turn, overrules the first strategy, is known to maintain bio-diversity, drive competition between bacterial strains, and preserve cooperation in social dilemmas. Here, we present a novel route to cyclic dominance in voluntary social dilemmas by adding to the traditional mix of cooperators, defectors, and loners, a fourth player type, risk-averse hedgers, who enact tit-for-tat upon paying a hedging cost to avoid being exploited. When this cost is sufficiently small, cooperators, defectors, and hedgers enter a loop of cyclic dominance that preserves cooperation even under the most adverse conditions. In contrast, when the hedging cost is large, hedgers disappear, consequently reverting to the traditional interplay of cooperators, defectors, and loners. In the interim region of hedging costs, complex evolutionary dynamics ensues, prompting transitions between states with two, three, or four competing strategies. Our results thus reveal that voluntary participation is but one pathway to sustained cooperation via cyclic dominance.
\end{abstract}

\maketitle

\section{Introduction}
\label{intro}

Large-scale cooperation is a basis for solving key societal problems including climate inaction \cite{vasconcelos_ncc13, pacheco_plrev14}, resource overexploitation \cite{hardin_g_s68, ciriacy1975common, bringezu2016multi}, imperfect vaccination \cite{wu_b_pone11, chen_prsb19, wang_z_pr16, kuga2019vaccinate}, antibiotic overuse \cite{chen_fp18}, crime occurrence \cite{orsogna_plr15}, and epidemic outbreaks \cite{helbing_jsp15, pastor_rmp15}. Cooperativeness as an altruistic act \cite{West2007Social} entails a cost for the actor in order for the recipient to enjoy a benefit. Cooperators are therefore fundamentally challenged by the most basic principles of Darwinian evolution, i.e., why should anyone act selflessly if only the fittest succeed? This puzzle has mobilized an unprecedented spectrum of researchers across disciplines to seek out mechanisms that sustain and/or promote cooperation \cite{West2007Social, pennisi_s05, nowak_s06, rand_tcs13, kraft_cobs15, perc_pr17, capraro_fp18, WestEvolutionary}.

Social dilemmas, a construct of evolutionary game theory \cite{hofbauer_98, nowak_06, sigmund_10}, capture the essence of the cooperation puzzle by contrasting individual and collective interests \cite{santos_pnas06, wang_plr15, allen2017evolutionary, fotouhi_rsif19}. An important subclass of social dilemmas is the voluntary social dilemma \cite{szabo_pre02b, wu_zx_pre05}, in which loners shape the evolutionary dynamics alongside traditional cooperators and defectors. Loners are risk-averse, and to avoid getting exploited they resort to an exceedingly simple strategy that generates a small fixed income regardless of what their opponents do. This prompts the emergence of cyclic dominance, whereby cooperators dominate loners who dominate defectors who dominate cooperators, thus sustaining cooperation while strategy abundances keep oscillating \cite{semmann_n03}. Cyclic dominance emerges elsewhere too, e.g., in the public goods game with correlated punishment and reward \cite{szolnoki_prx13}, in the ultimatum game with discrete strategies \cite{szolnoki_prl12}, in social dilemmas with jokers \cite{requejo_pre12b}, and by means of co-evolution \cite{szolnoki_epl09, perc_bs10}.

\begin{figure*}[t]
\centering
\includegraphics{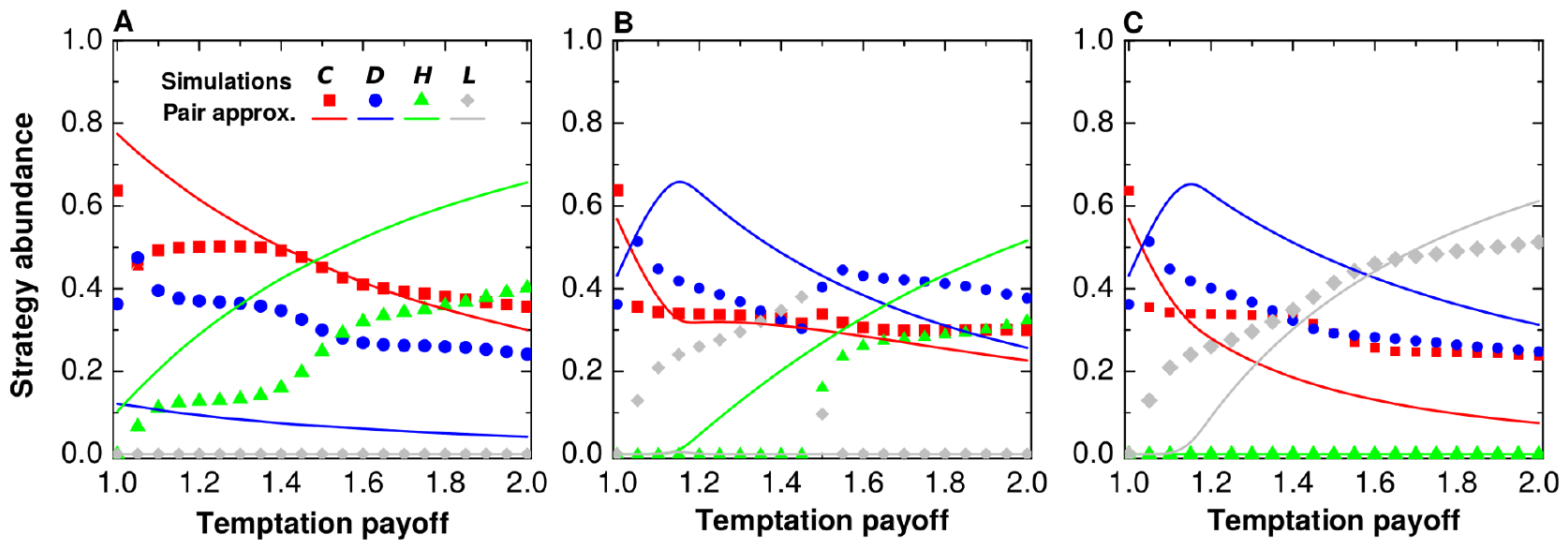}
\caption{Hedgers and loners interchangeably support cooperation via cyclic dominance. \textbf{A,} When hedging costs are low, $\alpha=0.05$, hedgers are more competitive than loners and manage to support cooperation via cyclic dominance over the full range of temptation payoffs. \textbf{B,} When hedging costs are intermediate, $\alpha=0.22$, hedgers are more competitive than loners only over a limited range of temptation payoffs. Cooperation is, nonetheless, supported via cyclic dominance as loners assume an analogous role to hedgers while the temptation payoff stays relatively low. \textbf{C,} With high hedging costs, $\alpha=0.4$, hedgers are no longer competitive at all. Cooperation is then maintained via cyclic dominance solely by loners. Simulation and pair-approximation results differ quantitatively, but they tell the same story qualitatively.}
\label{f01}
\end{figure*}

Cyclic dominance is often employed to study biodiversity \cite{frachebourg_prl96, frean_prsb01, kerr_n02, kirkup_n04, mobilia_pre06, reichenbach_n07, mobilia_jtb10, szolnoki_jrsif14, kelsic_n15} and competition in microbial populations \cite{durrett_jtb97, kirkup_n04, nahum_pnas11}. The subject of cyclic dominance is reviewed in detail in Ref. \cite{szolnoki_jrsif14}. Interestingly, when finite-size effects are taken into account, cyclic dominance may fail to sustain biodiversity and may even be responsible for extinction \cite{reichenbach_pre06}. Further results indicate that competition is key for sustaining biodiversity whilst ecosystem-wide patterns form by means of cyclical interactions \cite{jiang_ll_pre11}. Dynamically generated cyclic dominance, apart from sustaining biodiversity, offers alternative ways to sustain cooperators even in the face of large temptation to defect \cite{szolnoki_pre10b}. Considering more than three species in ecosystems, the existence of a phase with global oscillations, especially if the interaction graph contains multiple subloops and local cycles, prompts the conjecture that global oscillations are a general characteristic of realistic food webs \cite{rulquin_pre14}. Meanwhile, considering more than three strategies in evolutionary games shows that cyclic dominance emerges as an unexpected escape route from the adverse effects of antisocial punishment, also providing an explanation as to why second-order free-riding may not impede the evolutionary stability of punishment \cite{szolnoki_prx17}. Here, we considerably expand the scope of cyclic dominance in the generic and broadly adaptable setting of the voluntary social dilemma. Motivated by the ubiquity of risk aversion in the real world, we introduce a fourth player type, hedgers, who are also risk-averse, but in a more sophisticated and productive way than loners. Instead of avoiding meaningful interaction altogether, hedgers are willing to pay a hedging cost that allows them to learn the strategy of their opponents and thus avoid getting exploited. Specifically, a hedger enacts tit-for-tat play, but without cooperation in the first move \cite{axelrod_84, nowak_n93, szolnoki_pre09b}. If the opponent defects, the hedger also defects, whereas if the opponent cooperates, the hedger also cooperates.

In the described setting, hedgers may replace loners in a closed loop of dominance such that defectors invade cooperators, but cooperators invade hedgers, and hedgers invade defectors. We thus report a novel route to cyclic dominance in the voluntary social dilemma made possible by risk aversion. Strategy abundances again oscillate in time, but oscillations are now maintained via more elaborate and productive means than the simplistic loner strategy \cite{semmann_n03}. This is possible when the cost of hedging is sufficiently low. When the cost of hedging gets high, loners dominate hedgers, and we recover the original voluntary dilemma. For intermediate hedging costs, other solutions are also stable, including a four-strategy state in which all competing strategies coexist, or a two-strategy state in which cooperators and defectors coexist as is common in spatial prisoner's dilemma \cite{nowak_n92b}.

We hereafter proceed by detailing the mathematical model, describing the main results, and discussing their broader implications. We envision more direct applications to complex social systems, yet adaptations to biological systems also seem plausible.

\section{Mathematical Model}

\label{netw}
\begin{table}[b]
\centering
\setlength{\tabcolsep}{5mm}
\caption{\label{t00}Bilateral payoff matrix in the basic prisoner's dilemma game. A way to interpret this matrix is to assume that the payoff in a particular row and column is earned when a strategy in the first row is met with a strategy in the first column. Thus, payoff $S$ is earned by a cooperator $C$ meeting a defector $D$. The payoff order is $T>R>P>S$ to encourage defection and discourage cooperation. Also, $2R>T+S$ makes mutual cooperation ($2R$) more beneficial for the collective than defecting against a cooperator ($T+S$).}
\begin{tabular}{r|ccccc}
    & $C$        & $D$    \\
\hline
$C$ & $R$        & $S$    \\
$D$ & $T$        & $P$    \\
\end{tabular}
\end{table}

Our model is a variant of the prisoner's dilemma game (PDG) extended to incorporate loners and hedgers. In a common PDG \cite{axelrod_84}, mutual cooperation is rewarded by the reward payoff, $R$, whereas mutual defection is punished by the punishment payoff, $P$. Furthermore, defection is encouraged by the largest payoff, temptation $T$, while cooperation is discouraged by the lowest sucker's payoff, $S$. This is because of the payoff ranking $T>R>P>S$, and the fact that $T$ is earned by a defector upon meeting a cooperator who then earns $S$. In the repeated version of the game, it is additionally assumed that $2R>T+S$ \cite{szabo_pr07}. The corresponding bilateral payoff matrix is given in Table~\ref{t00}. Although the game seems to be defined by four payoffs, there is a well known rescaling to two parameters that leaves the evolutionary dynamics unaffected \cite{wang_plr15}. For even more simplicity, studies oftentimes resort to the weak PDG \cite{nowak_n92b}, in which only one parameter, $T=b$, controls how strong the dilemma is in the sense that the larger the value of $b$, the stronger is the temptation to defect. The other payoffs are simplified to $R=1$ and $P=S=0$.

We extended the weak PDG to include loners and hedgers. A loner avoids complications by always earning a small payoff, $\sigma{\ll}b$, which is also earned by anyone who meets the loner. A hedger, by contrast, is much more sophisticated. Due to the hedger's risk aversion, they pay a hedging cost, $\alpha$, to learn the opponent's strategy, and then enact tit-for-tat play by cooperating with a cooperator or defecting against a defector. With the payoffs of all four strategies defined (Table~\ref{t01}), it is worthwhile to briefly examine implications for the evolutionary dynamics.

\begin{table}[b]
\centering
\setlength{\tabcolsep}{5mm}
\caption{\label{t01}Bilateral payoff matrix of the four-strategy model. As in Table~\ref{t00}, the payoff in a particular row and column is earned when a strategy in the first row is met with a strategy in the first column. For example, payoff $b$ is earned by a defector $D$ meeting a cooperator $C$, whereas payoff $\sigma$ is earned by a hedger $H$ meeting a loner $L$.}
\begin{tabular}{r|ccccc}
    & $C$        & $D$       & $L$      & $H$        \\
\hline
$C$ & 1          & 0         & $\sigma$ & 1          \\
$D$ & $b$        & 0         & $\sigma$ & 0          \\
$L$ & $\sigma$   & $\sigma$  & $\sigma$ & $\sigma$   \\
$H$ & 1-$\alpha$ & -$\alpha$ & $\sigma$ & 1-$\alpha$ \\
\end{tabular}
\end{table}

\begin{figure*}[t]
\centering
\includegraphics[scale=1.0]{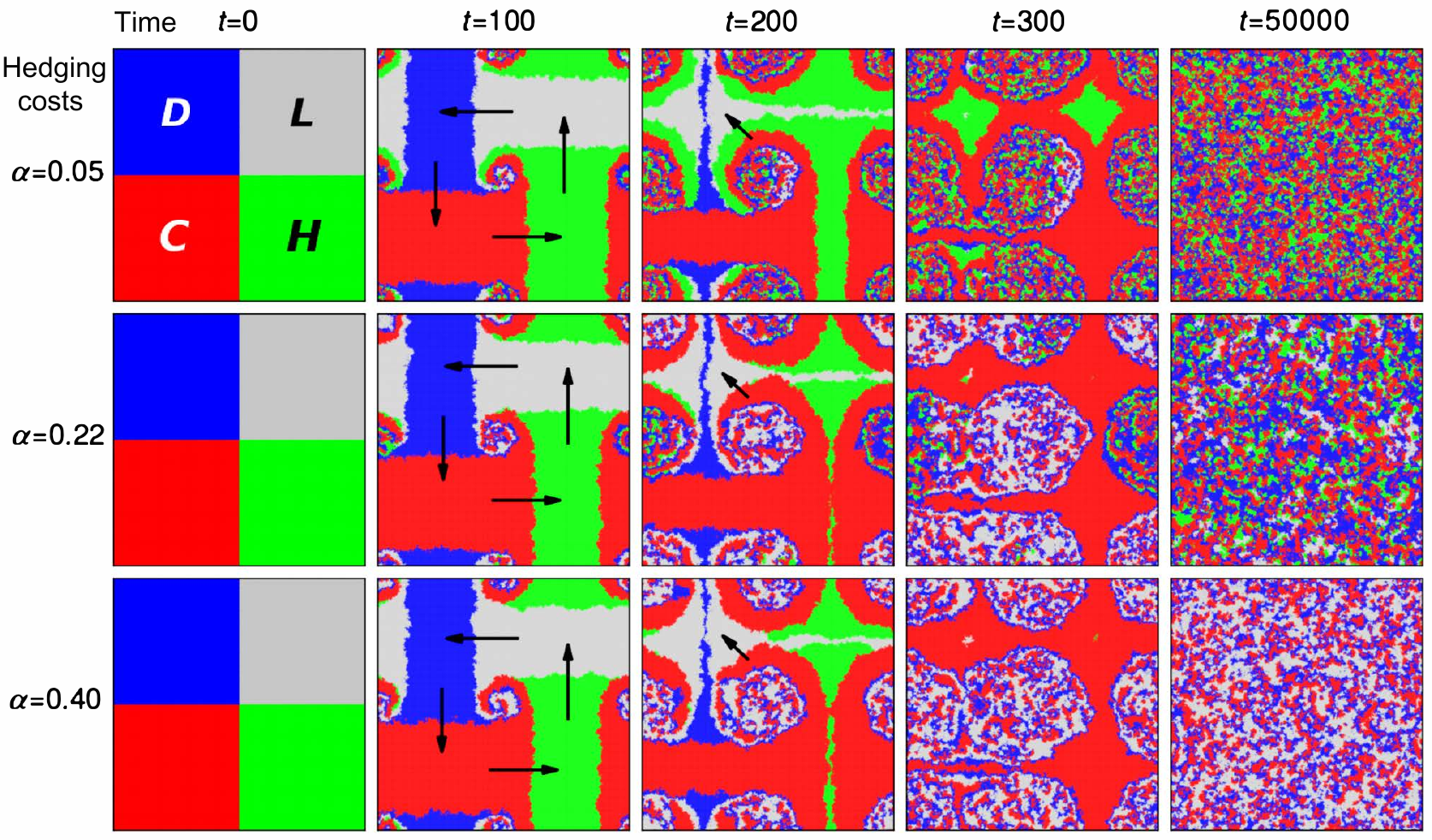}
\caption{Hedging costs shape the fate of hedgers. Shown are the snapshots of evolutionary dynamics at different time steps (columns) and for different hedging costs (rows). Early in the evolutionary dynamics, defectors invade cooperators who invade hedgers who invade loners. Later, however, it is hedging costs that determine the fate of hedgers. When these costs are low (upper row), hedgers pay little for denying defectors the temptation payoff, as well as enjoying the reward payoff together with cooperators. All this helps hedgers to outlive loners who eventually disappear. When hedging costs are high (bottom row), risk aversion is a burden that cannot be offset by the reward payoff. Hedgers eventually disappear. Finally, when hedging costs are intermediate (middle row), there is a balance of power of sorts between hedgers and loners, and all four strategies survive. The irregular mixing of strategies after transient evolutionary dynamics is a signature of cyclic dominance.}
\label{f02}
\end{figure*}

Focusing momentarily on the $CDL$ combination of players \cite{szabo_pre02b, wu_zx_pre05}, we see that in the absence of cooperators, loners receive the same or larger payoff than defectors, giving the former strategy a decisive evolutionary advantage over the latter. The situation is similar for defectors against cooperators in the absence of loners, and for cooperators against loners in the absence of defectors. Based on these observations, it is a fair guess even before running any simulations that cyclic dominance may emerge, especially when the game is embedded into a spatial structure.

For the $CDH$ combination of players, the setup is somewhat different. The relation between cooperators and defectors in the absence of hedgers remains unchanged, of course, and the relation between cooperators and hedgers mimics that of cooperators and loners because, in the absence of defectors, it is cooperators who always attain a larger payoff. The relation between defectors and hedgers is complicated by the fact that, in the absence of cooperators, either strategy may earn more depending on who interacts with whom and what the exact value of hedging cost $\alpha$ is. Nonetheless, when the cost is small and hedgers sufficiently abundant, they should be able to overcome defection, possibly giving rise to cyclic dominance once again. We examined this supposition using Monte Carlo simulations \cite{szabo_pr07, hauert_ijbc02} as well as the pair approximation technique \cite{szabo_pr07, hauert_ajp05} (Supplementary information). Monte Carlo simulations of evolutionary games are an individual-based approach in which actions of each individual are implemented explicitly to generate a payoff that is then compared to the payoffs of other individuals, followed by imitation of those who performed better. Simulations are typically run until the system reaches a stable state as indicated by an order parameter, e.g., the fraction of cooperators. The pair approximation method provides alternative means of tracking the system's state. Specifically, by focusing on the frequencies and the proportions of strategy pairs, the method yields a set of differential equations that should implicitly encode the behaviors that are explicit in Monte Carlo simulations. As such, pair approximation can be used to verify the results of Monte Carlo simulations or offer a different perspective on particular outcomes.

For the purpose of Monte Carlo simulations, we arranged a total of $N=L^2$ agents to form a square lattice of length $L$ with periodic boundary conditions. Each agent is characterized by a strategy vector, $S_x=(p_x^C,p_x^D,p_x^L,p_x^H)^T$, where $p_x^*$ is the probability with which agent $x$ behaves as a cooperator, defector, loner, or hedger. A direct consequence is that $p_x^C+p_x^D+p_x^L+p_x^H=1$. We additionally imposed that exactly one $p_x^*$ equals unity, meaning that the only four possible strategy vectors for agent $x$ are $S_x=(1,0,0,0)^T{\equiv}C$, $S_x=(0,1,0,0)^T{\equiv}D$, $S_x=(0,0,1,0)^T{\equiv}L$, and $S_x=(0,0,0,1)^T{\equiv}H$. Then the payoff earned by this agent in a single round of the game is
\begin{equation}
P_x=\sum_{y\in\mathrm{N}_x}{{S_x}^T\mathbf{M}S_y},
\label{eq1}
\end{equation}
where $\mathrm{N}_x$ is the set of agent $x$'s neighbors, $\mathbf{M}$ is the matrix of elementary payoffs as displayed in Table~\ref{t01}, and $S_y$ is the strategy vector of $x$'s neighbor $y$. Depending on the earned payoffs, agents update their strategies using the Fermi rule \cite{szabo_pr07, hauert_ijbc02}, which ensures that poorer performing players eventually adopt the strategies of their better performing counterparts. Specifically, agent $x$ imitates the strategy of a randomly chosen neighbor $y\in\mathrm{N}_x$ with probability
\begin{equation}
W_{S_x{\leftarrow}S_y}=\frac{1}{1+\exp{\left(\frac{P_x-P_y}{K}\right)}},
\label{eq2}
\end{equation}
where $K$ quantifies the irrationality of selection, i.e., the larger the $K$ is, the greater the probability that agent $x$ imitates agent $y$ even if $P_x>P_y$.

To run simulations from this point on, we only needed to specify parameter values. The payoff parameter whose value was held constant throughout this study is $\sigma=0.3$. To uphold constraint $2R>T+S$, we examined temptation payoffs in the range $1{\leq}b{\leq}2$; $b$=2 thus signifies the strongest dilemma. Furthermore, we presented the results for hedging costs in the range $0{\leq}\alpha{\leq}0.4$ because there are no qualitative changes in the system's behavior for $\alpha>0.4$. The lattice size systematically varied in the range $200{\leq}L{\leq}1500$ to make sure that the results are not due to the finite size effects. The closed neighborhood was von Neumann's \cite{hauert_ijbc02} throughout the study. Relative to an arbitrarily selected focal agent, such a neighborhood includes the closest 'north', 'east', 'south', and 'west' agents, implying a constant node degree of $k=4$. Finally, the irrationality of selection parameter was kept constant at $K=0.1$. We ran each simulation over 50\,000 time steps to guarantee that the transient dynamics had passed.

\section{Results}

\begin{figure}
\centering
\includegraphics{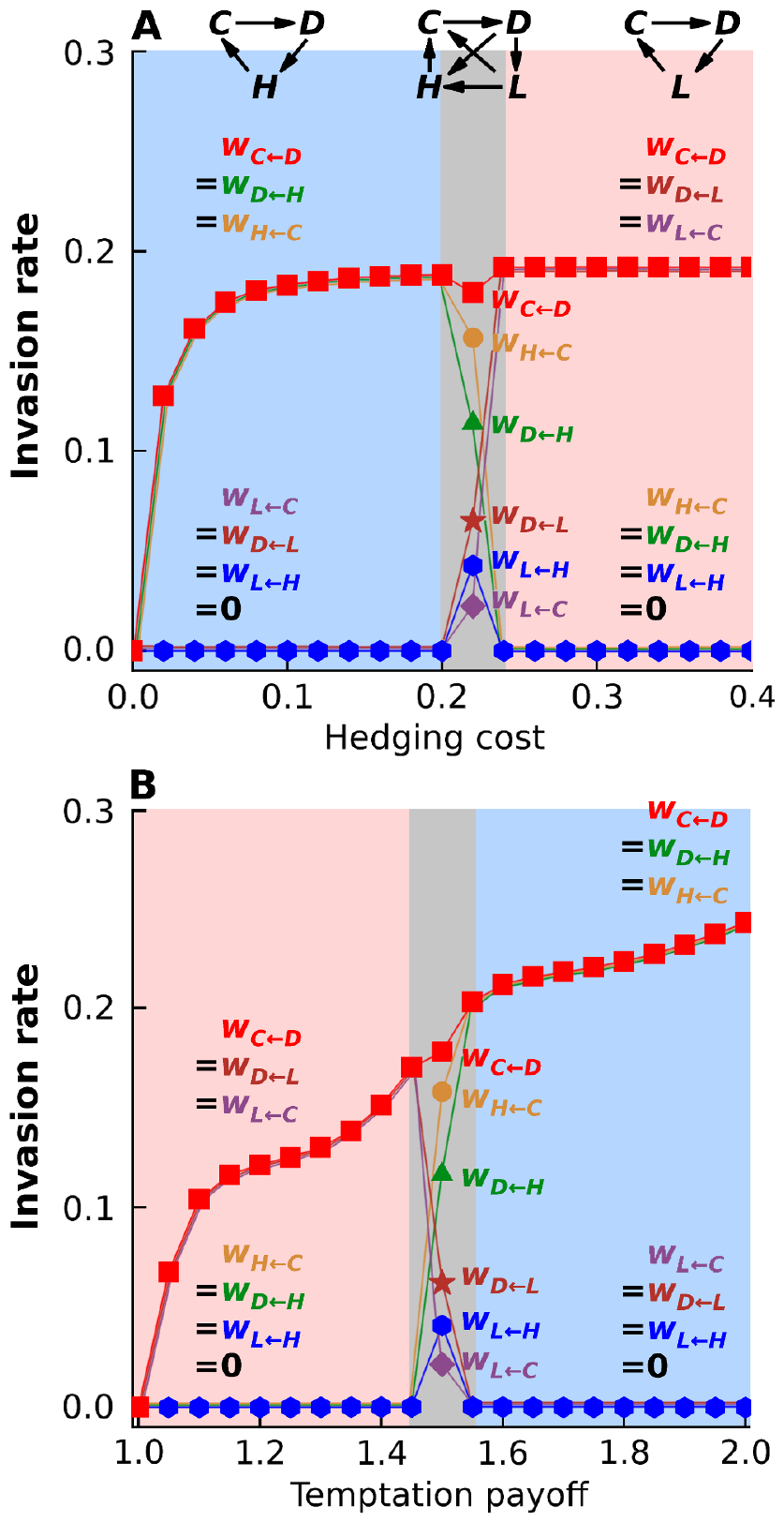}
\caption{Average invasion rates reveal cyclic dominance in action. \textbf{A,} Fixing the temptation payoff to $b=1.5$, we plotted average invasion rates as the functions of hedging costs, $\alpha$. For small $\alpha$ values, cyclic dominance between cooperators, defectors, and hedgers (light-blue area) is reflected in the fact that defectors invade cooperators who invade hedgers who invade defectors all with the same average invasion rate. For large $\alpha$ values, the situation is analogous, except that loners take the place of hedgers (light-red area). For a narrow window of intermediate $\alpha$ values, the two cyclic dominance loops intermix, with hedgers actually invading loners with a small invasion rate ($w_{L{\leftarrow}H}$) that is more than compensated for by loners invading defectors ($w_{D{\leftarrow}L}>w_{L{\leftarrow}H}$), which enables all four strategies to coexist. \textbf{B,} Similar results are obtained by fixing the hedging cost to an intermediate value of $\alpha=0.22$, and then varying the temptation benefit. Here, cyclic dominance with loners emerges for small $b$ values (light-red area), whereas cyclic dominance with hedgers emerges for large $b$ value (light-blue area).}
\label{f03}
\end{figure}

Simulations reveal that hedgers are able to support cooperation via cyclic dominance. When the hedging cost is sufficiently low, hedgers outlive loners and remain abundant over the whole range of temptation payoffs (Fig.~\ref{f01}A). In doing so, hedgers enter into a loop of cyclic dominance with cooperators and defectors, helping the former to avoid being overrun by the latter. How this transpires is more clearly seen in the snapshots of evolutionary dynamics in which defectors invade cooperators who invade hedgers who invade loners early on (second column in Fig.~\ref{f02}). Afterwards, however, hedgers maintain their presence by paying little cost while denying defectors the temptation payoff and enjoying the reward payoff with cooperators (top row in Fig.~\ref{f02}). Under such conditions, loners are the least competitive player type whose eventual disappearance marks the beginning of cyclic dominance between cooperators, defectors, and hedgers. That these player types indeed cyclically dominate one another is reflected in the irregular mixing pattern of agents post transient evolutionary dynamics (fifth column in Fig.~\ref{f02}).

Loners start replacing hedgers as the hedging cost increases. This first happens for small (Fig.~\ref{f01}B) and then all (Fig.~\ref{f01}C) values of the temptation payoff. Despite losing the support from hedgers, cooperators do not succumb to defectors because it is now loners who enter the loop of cyclic dominance. The snapshots of evolutionary dynamics illustrate this turn of events. As hedgers become overburdened by their risk-averse strategy due to the high hedging cost, they become the least competitive player type and eventually die out (bottom row in Fig.~\ref{f02}). The recognizable irregular mixing pattern of agents engaged in cyclic dominance resurfaces again after transient evolutionary dynamics, but now with loners having taken the place of hedgers. Interestingly, although hedgers and loners almost entirely exclude one another, there is a narrow window of hedging costs and temptation payoffs in which these two player types coexist alongside cooperators and defectors (middle row in Fig.~\ref{f02}).

To glimpse into the heart of cyclic dominance, we estimated invasion rates post transient evolutionary dynamics. If strategy $S_x$ had been invaded by strategy $S_y$, we defined the corresponding invasion rate, $w_{S_x{\leftarrow}S_y}>0$, as the net positive fraction of transitions from strategy $S_x$ to strategy $S_y$, mediated by the Fermi rule in a single Monte Carlo time step. With four player types, six invasions had been possible. These happened to be $w_{C{\leftarrow}D}$, $w_{L{\leftarrow}C}$, $w_{H{\leftarrow}C}$, $w_{D{\leftarrow}L}$, $w_{D{\leftarrow}H}$, and $w_{L{\leftarrow}H}$.

\begin{figure}
\centering
\includegraphics[scale=1.0]{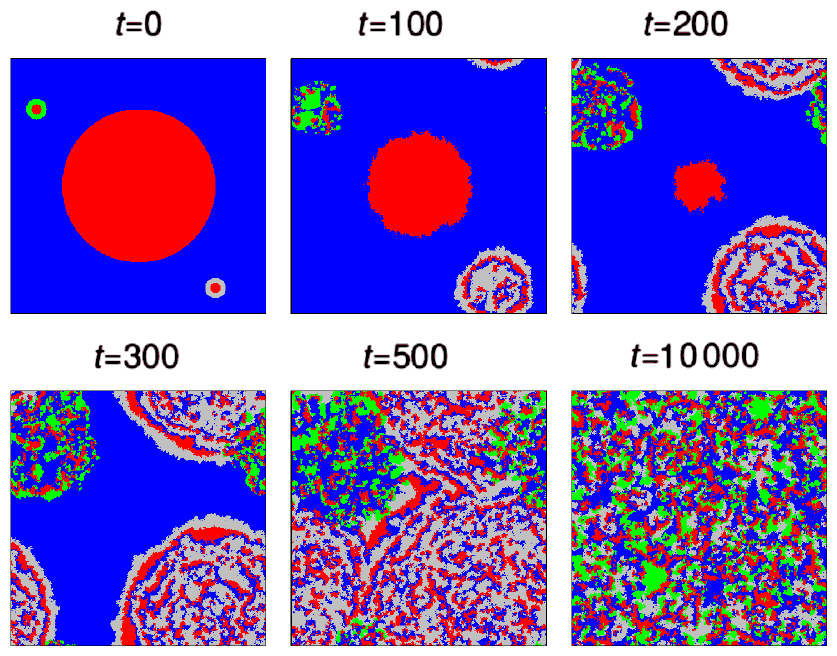}
\caption{Hedgers and loners protect cooperators even in the harshest of conditions. Shown is the time evolution of strategies such that, initially, there is a one giant cluster of cooperators (red) surrounded by defectors (blue), and two smaller clusters of cooperators separated from defectors by a thin layer of hedgers (green) in one case and loners (gray) in the other. The temptation payoff, $b=2.0$, is at the maximum of the prisoner's dilemma limits. The hedging cost of $\alpha=0.32$ is intermediate. Under such conditions, the giant cooperative cluster quickly erodes under exploitation by defectors. Smaller cooperative clusters, by contrast, turn into mixes of cooperators, defectors, and loners or hedgers, that spread like ripples across a lattice, here, of size $L=500$.}
\label{f04}
\end{figure}

\begin{figure}
\centering
\includegraphics[scale=1.0]{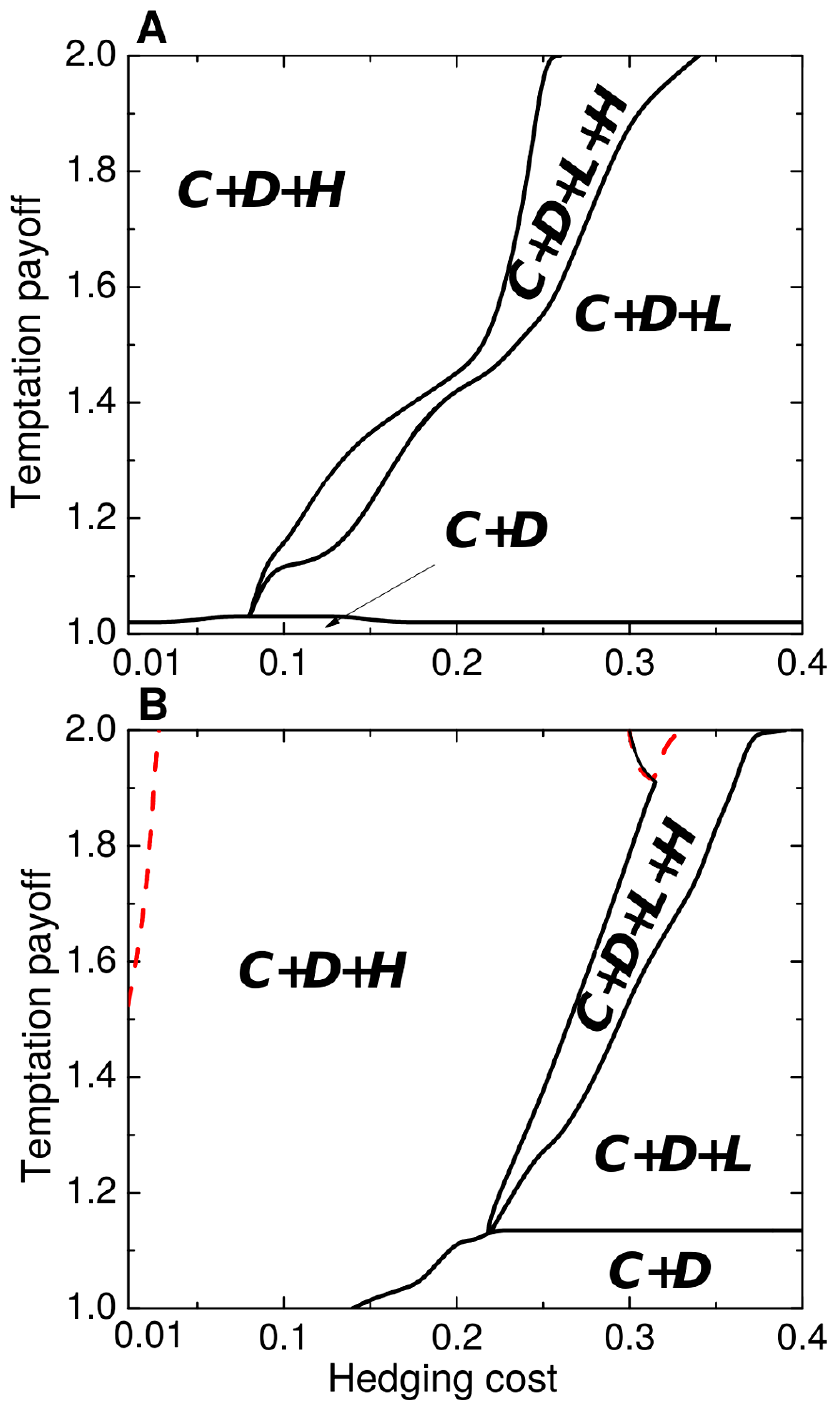}
\caption{Cyclic dominance is a widespread phenomenon. Shown are the phase diagrams due to Monte Carlo simulations and the pair approximation, revealing the model's dynamical regimes as function of the hedging cost and the temptation payoff. \textbf{A,} Cyclic dominance arises almost everywhere in the phase space of Monte Carlo simulations, with hedgers surviving at low and loners at high hedging costs. There is also a narrow domain in which all four strategies coexist. \textbf{B,} Pair approximation yields a qualitatively similar distribution of strategy prevalences. Dashed red curves delineate subdomains where the system converges to limit cycles by which the abundances of surviving player types oscillate in time (see Fig.~\ref{f06}).}
\label{f05}
\end{figure}

Invasion rates clearly demonstrate cyclic dominance in action. For an intermediate temptation payoff, $b=1.5$, and relatively low hedging costs, cooperators, defectors, and hedgers get locked in a cyclic dominance loop as evidenced by the equality of invasion rates $w_{C{\leftarrow}D}=w_{D{\leftarrow}H}=w_{H{\leftarrow}C}$ (Fig.~\ref{f03}A). All invasion rates involving loners equal zero because players of this type get outcompeted. As hedging costs increase, the fortunes of hedgers and loners reverse, and now the latter get locked in the cyclic dominance loop with cooperators and defectors as indicated by equalities $w_{C{\leftarrow}D}=w_{D{\leftarrow}L}=w_{L{\leftarrow}C}$ (Fig.~\ref{f03}A). All invasion rates involving hedgers equal zero because hedgers get outcompeted, just as loners did in the opposite situation. All four player types coexists only over a narrow window of intermediate hedging costs. Here, cooperators are drained in favor of defectors, but replenished at the expense of hedgers and loners, i.e., $w_{C{\leftarrow}D}=w_{H{\leftarrow}C}+w_{L{\leftarrow}C}$. Defectors are drained in favor of hedgers and loners, but replenished at the expense of cooperators, i.e., $w_{D{\leftarrow}H}+w_{D{\leftarrow}L}=w_{C{\leftarrow}D}$. Loners are drained in favor of hedgers and cooperators, but replenished at the expense of defectors, i.e., $w_{L{\leftarrow}H}+w_{L{\leftarrow}C}=w_{D{\leftarrow}L}$. Finally, hedgers are drained in favor of cooperators, but replenished at the expense of defectors and loners, i.e., $w_{H{\leftarrow}C}=w_{D{\leftarrow}H}+w_{L{\leftarrow}H}$. Completely analogous results hold when the hedging cost is intermediate, $\alpha=0.22$, and the temptation payoff increases in the range $1{\leq}b{\leq}2$ (Fig.~\ref{f03}B).

\begin{figure*}
\centering
\includegraphics[width=\linewidth]{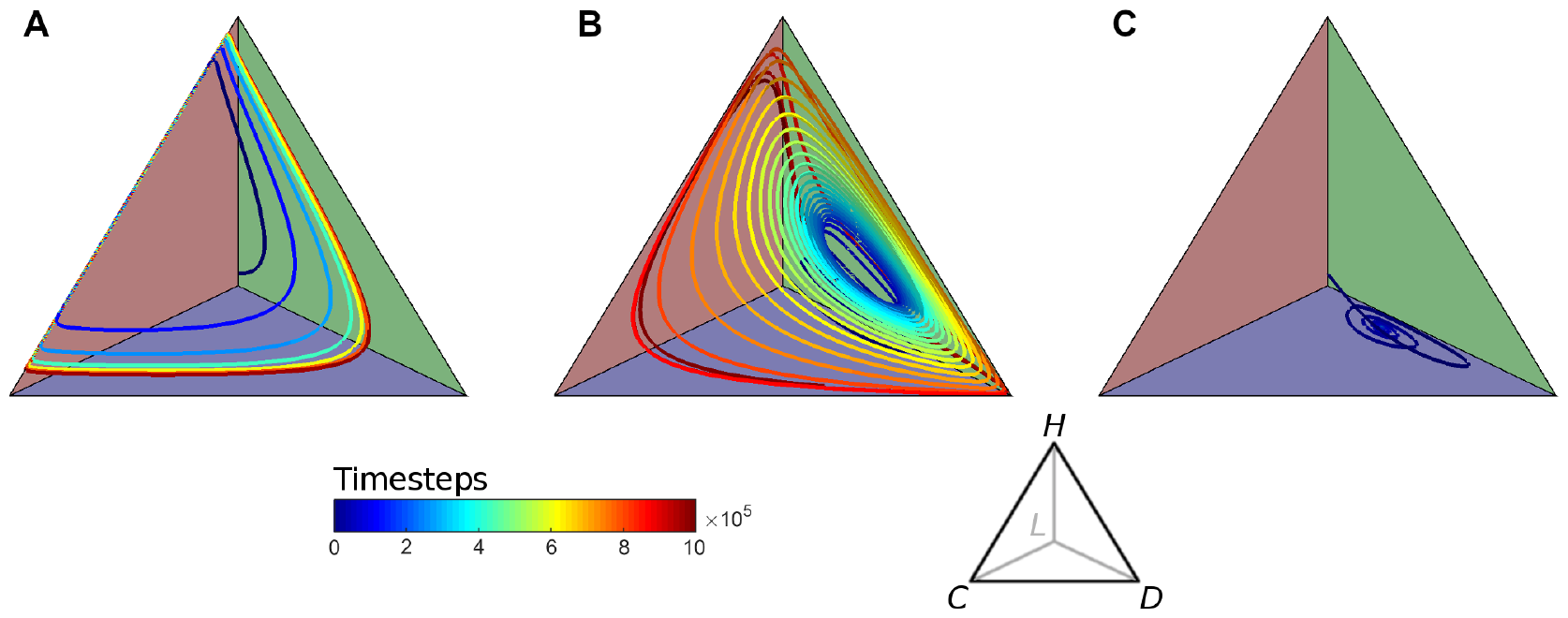}
\caption{Exceeding simplicity of the loner strategy is manifest in the absence of an explicit spatial structure. Specifically, when approximating the system using pair approximation, hedgers can still maintain cyclic dominance such that strategy abundances oscillate in time, but loners cannot. \textbf{A,} With low hedging costs, $\alpha=0.02$, and a sufficiently large temptation payoff, $b=2.0$, loners quickly disappear as the system converges to a planar limit cycle with positive cooperator, defector, and hedger abundances. The limit cycle is planar because the three strategy abundances must sum to unity. \textbf{B,} When the hedging cost increases to $\alpha=0.32$, all four strategies coexist and the system converges to a three-dimensional limit cycle. The limit cycle is three-dimensional because strategy abundances must again sum to unity. \textbf{C,} With high hedging costs, $\alpha=0.40$, it is hedgers, and with them oscillations in strategy abundances, who quickly disappear. The system thus approaches a stable equilibrium of cooperator, defector, and loner strategies. The colorbar indicates time steps.}
\label{f06}
\end{figure*}

Cyclic dominance emerges rather independently of the initial conditions and even in the face of the strongest dilemma, $b=2.0$. To demonstrate this, we present simulations of the evolutionary dynamics starting from one large cluster of cooperators surrounded entirely by defectors, and two small clusters of cooperators, but separated from defectors by a thin layer of hedgers in one case and loners in the other (Fig.~\ref{f04}). Setting the hedging cost to an intermediate value, $\alpha=0.32$, we find that defectors quickly chip away at the large cooperative cluster, while smaller cooperative clusters turn into a mix of cooperators, defectors, and hedgers or loners, and spread in a ripple-like manner through the lattice, until cyclic dominance becomes omnipresent (Fig.~\ref{f04}).

To fully understand evolutionary dynamics as a function of the temptation payoff and the hedging cost, we mapped the system's phase space. We found four distinct phases: cooperators and defectors only ($C+D$), cooperators-defectors-hedgers ($C+D+H$), cooperators-defectors-loners ($C+D+L$), and all four player types ($C+D+L+H$). When the temptation benefit is very small, the dilemma is minimal and the system evolves to the $C+D$ phase irrespective of the hedging cost (Fig.~\ref{f05}A). As the temptation payoff increases, it is the hedging cost that largely determines how the system evolves. For relatively small hedging costs, the $C+D+H$ phase is prevalent, whereas for relatively large hedging costs, the $C+D+L$ phase is prevalent (Fig.~\ref{f05}A). There is only a narrow window of intermediate hedging costs in which the system evolves to the $C+D+L+H$ phase (Fig.~\ref{f05}A), showing again that hedgers and loners almost interchangeably support cooperation through cyclic dominance. The pair approximation yields a qualitatively similar phase diagram, despite some obvious quantitative differences relative to Monte Carlo simulations (Fig.~\ref{f05}B).

It is not entirely equivalent whether cyclic dominance is established by hedgers or loners. Namely, in the spatial voluntary dilemma, cyclic dominance gives rise to oscillating strategy abundances everywhere on the $b$--$\alpha$ phase diagram, but without spatial structure only cyclic dominance involving hedgers can do so. Fixing the temptation payoff to $b=2.0$ and choosing a low hedging cost of $\alpha=0.02$ places the system in the $C+D+H$ phase such that the pair approximation generates an open cycle trajectory that converges to a planar limit cycle (Fig.~\ref{f06}A). The limit cycle is planar because loners die out and the sum of strategy abundances for the remaining three player types is constrained to unity. Increasing the cost of hedging to a higher value of $\alpha=0.32$ sets the system into the $C+D+L+H$ phase in which evolutionary dynamics converges to a three-dimensional limit cycle (Fig.~\ref{f06}B) because now all four player types co-exist, but their abundances are still constrained to unity. Finally, with a high hedging cost of $\alpha=0.4$, the system is pushed into the $C+D+L$ phase, where the pair approximation yields a trajectory with small open cycles that disappear as the system converges to a stable equilibrium with zero hedgers (Fig.~\ref{f06}C). A degree of cooperation is still maintained in such an equilibrium, but cyclic dominance as the underlying cause stays masked by the static outcome of evolutionary dynamics.

\section{Discussion}

Herein, we revealed a novel route to cyclic dominance in voluntary social dilemmas brought about by risk averse players called hedgers. These players outwit defectors by paying a hedging cost to avoid getting exploited. In the presence of cooperators, however, risk aversion is redundant and thus a burden that hampers performance. Hedgers accordingly invade defectors, defectors---as usual in social dilemmas---invade cooperators, and cooperators invade hedgers to close the loop of cyclic dominance. This effectively sustains cooperation even under adverse conditions, when the temptation to defect is high. Moreover, this shows that cooperativeness by way of cyclic dominance can emerge due to more elaborate and productive means than the exceedingly simple loner strategy.

Hedgers may superficially resemble a simple hybrid of cooperators and defectors, yet as risk-averse conditional cooperators they are much more. Plain conditional cooperators, due to their relationship to the tit-for-tat strategy \cite{fehr2004social}, have played a central role in the studies on the evolution of cooperation \cite{frey2004social, rustagi2010conditional, suzuki2010neural}. Adding to this the fact that risk aversion is ubiquitous in real life \cite{halek2001demography}, the importance of considering risk-averse conditional cooperators can hardly be overstated.

Our research has important implications for promoting and sustaining cooperation in human societies. An increasing realism via higher strategic complexity \cite{capraro_prsb15, hauser_n14, jordan_n16} is necessarily accompanied with an increasing complexity of evolutionary outcomes as they become dependent not just on individual strategic relations prescribed by payoff elements, but also on dynamical relations and spontaneously emerging alliances between strategy subsets \cite{szolnoki_jrsif14}. Cyclic dominance with hedgers is one such example, but we expect to see many more in the near future. This, in turn, predicts considerable challenges in steering complex social systems towards a desired cooperative state. Simple interventions, as frequently enacted by decision makers, often turn out to be naive and risky propositions precisely because of a limited understanding of the underlying spatio-temporal dynamics that govern social complexity. Relevant examples include producing and appointing inadequately trained personnel~\cite{jacobson1998training}, flawed models and agency costs associated with those models~\cite{gerding2009code}, misunderstanding or ignoring systemic risks~\cite{haldane2011systemic}, etc.

Aside from applications to social dynamics, it is possible to envision analogous, if more complicated, setups in biological systems as well. Living tissues are genetically identical cell structures of the same differentiation fate that work cooperatively to support homeostasis~\cite{celiker2013cellular}. Such cooperative cell structures are susceptible to cheating~\cite{celiker2013cellular}, i.e., individual cells occasionally defect and turn malignant by either increasing their own fitness or decreasing the fitness of the surrounding cells~\cite{merlo2006cancer, bozic2010accumulation, greaves2012clonal}. The system is, furthermore, policed by free-ranging immune cells that scan for malignant activity~\cite{reiche2004stress, finn2012immuno}, and actively punish defectors at a cost~\cite{derting2003immune, hanssen2004costs}, while receiving a compensation for that cost through interactions with cooperative cells. In this sense, cooperative cells would not ``invade'', but rather ``modulate'' the activity of immune cells, yet the resulting dynamics should still resemble cyclic dominance. Understanding such dynamics by, for example, mapping the system's phase space, could help to devise novel treatments for controlling malignant activity~\cite{finn2012immuno, couzinfrankel2013cancer}.

While there is a potential for widespread uses, we hope that our research will, at the very least, lead to better understanding of cooperation in social dilemmas. In this context, a particularly promising way forward is to pair up theoretical treatises of evolutionary game dynamics with laboratory \cite{rand2012promise, wang2017onymity} and even field experiments. Doing so should put theory to the test, help identify more solid fundamentals on which to build upon, and overall strongly aid the quest for a better and more sustainable future.

\paragraph*{Author Contributions.} H.G. and Z.W. conceived the research. H.G. performed simulations. All coauthors discussed and interpreted the results, and wrote the manuscript.
\paragraph*{Data Accessibility.} Programming code used to run simulations is publicly available at DOI 10.17605/OSF.IO/7CV69.
\paragraph*{Funding Statement.} This research was supported by the National Natural Science Foundation of China (Nos. 11931015 and U1803263), Key Area Research \& Development Program of Guangdong Province (No. 2019B010137004), the Fundamental Research Funds for the Central Universities (No. 3102019PJ006), Shannxi Natural Science Foundation Key Program (No. 2019ZDLGY17-07), the Slovenian Research Agency (Nos. J4-9302, J1-9112, and P1-0403), the Government of Arag\'on and the European Regional Development Fund (FEDER) funds (No. E36-17R to FENOL), Ministry of Economy and Competitiveness (MINECO) and FEDER funds (No. FIS2017-87519-P), and Intesa Sanpaolo Innovation Center. The funders had no role in study design, data collection, analysis, decision to publish, nor preparation of the manuscript.


%

\clearpage

\setcounter{page}{1}
\renewcommand\thepage{\roman{page}}

\setcounter{equation}{0}
\setcounter{figure}{0}
\setcounter{table}{0}
\makeatletter
\renewcommand{\theequation}{S\arabic{equation}}
\renewcommand{\thefigure}{S\arabic{figure}}
\renewcommand{\bibnumfmt}[1]{[S#1]}
\renewcommand{\citenumfont}[1]{S#1}

\onecolumngrid
\begin{center}
\begin{large}
\textbf{Supplementary information}
\end{large}
\end{center}

\twocolumngrid

\section*{Pair approximation}

Local spatial interactions on a lattice are often key to understanding dynamical processes. To capture important features of the local spatial behaviour, a commonly employed technique is pair approximation, which focuses on tracking the frequencies and the proportions of strategy pairs.

In the following, we derive equations that capture the dynamics of pair proportions in an evolutionary game with four strategies, $C$, $D$, $L$, and $H$, taking place on a large square lattice. To that end, we first define the space and the variables of interest, then we specify relationships that estimate configuration and transition probabilities, and finally, we list differential equations for pair proportions. Notation and derivations hereafter follow existing literature, with some additional details to make the exposition reader-friendly \citep{Perc2006, szabo_pr07, Lion2016}.

\paragraph*{\textbf{Space.}} In an evolutionary game with four strategies, every node of a network comprising $N$ nodes is in one of the four states, $C$, $D$, $L$, or $H$, respectively denoting cooperators, defectors, loners, and hedgers. Let $\Sigma=\{C,D,L,H\}$ be a set of states. Each node $i$ is in state $\omega_{i}$, $\omega_{i}\in\Sigma$. The state of the network is then $\bm{\omega}=\left(\omega_{i}\right)\in\Omega$, where $\Omega=\Sigma^N$ is the set of all possible network states. Vector functions $\bm{s}:\Omega\rightarrow \{0,1\}^N$, $\bm{s}=\bm{c},\bm{d},\bm{l},\bm{h}$, indicate network nodes in states $S=C,D,L,H$, respectively. In other words, $\bm{s}\left(\bm{\omega}\right)=\left(s_{i}\left(\bm{\omega}\right)\right)$, where
\begin{equation}
s_{i}\left(\bm{ \omega}\right)= \begin{cases}
  1  & \quad \text{if } \omega_i=S,\\
  0  & \quad \text{otherwise}.
\end{cases}
\end{equation}
Adjacency matrix $\bm{A}$, i.e., the matrix of links, is an $N{\times}N$ symmetrical matrix such that $A_{ij}=1$ if nodes $i$ and $j$ are linked, and $A_{ij}=0$ otherwise. No node is connected to itself, implying that $A_{ii}=0$. In the case of a large square lattice, each node is connected to $n=4$ neighbouring nodes, further implying that $\sum_{j=1}^n{A_{ij}}=4$.

\paragraph*{\textbf{Variables.}} Our goal is to trace the dynamics of a small set of variables well-defined on set $\Omega$. Variables containing information on global and local abundances of strategies are:
\\
\\
\textit{Number of nodes in state S}
\begin{equation}
\left[S\right]=\sum_{i=1}^{N}s_{i}=\bm{ s^T\,1}  \label{eqn:konc1}
\end{equation}
\textit{Number of $(S,S')$ strategy pairs}
\begin{equation}
\left[S:S'\right] = \sum_{i=1}^{N} \sum_{j=1}^{N} A_{ij} s_{i} s'_{j} =\bm{ s^T\,A\,s'} \label{eqn:konc2}
\end{equation}
\textit{Average number of $S$ neighbouring $S'$}
\begin{equation}
\left[S|S'\right]=\frac{\left[S:S'\right]}{\left[S'\right]} \label{eqn:konc3}
\end{equation}
From these definitions, it follows that \citep{Lion2016}
\begin{align}
\sum_{S}\left[S\right] &= N,  \label{eqn:konc4} \\
\sum_{S'}\left[S:S'\right] &= n\left[S\right],   \label{eqn:konc5} \\
\sum_{S''}\left[S:S':S''\right] &= \frac{n(n-1)}{N}\left[S\right]\left[S'\right],   \label{eqn:konc6}
\end{align}
where once again $N$ is the total number of nodes in the network, and $n$ is the average node degree. Note that Eq.~(\ref{eqn:konc4}) is exact, whereas Eqs.~(\ref{eqn:konc5} \& \ref{eqn:konc6}) hold exactly only for regular networks; for other random networks these equations hold in expectation. Furthermore, Eqs.~(\ref{eqn:konc4} \& \ref{eqn:konc5}) imply that the total number of pairs equals $nN$, or in the case of a large square lattice $4{\cdot}N$.

From Eq.~(\ref{eqn:konc2}), we can formally derive
\begin{equation}
\left[S:S'\right] =\bm{s^TAs'}= \bm{\left(s^TAs'\right)^T} = \bm{{s'}^TAs} = \left[S':S\right].
\end{equation}
This relationship prescribes how to count node pairs. Namely, the number of $\left(S,S'\right)$ pairs is given as follows: for each focal node $i$ in state $S$, count its neighbouring nodes $j$ in state $S'$ to obtain the total number of $\left(S,S'\right)$ pairs in the network. Furthermore, the symmetry of adjacency matrix $A_{ij}$ dictates that, if $S=S'$, a pair of neighbouring nodes $i$ and $j$ in state $\left(S,S\right)$ will be counted twice, once when $i$ is the focal node, and once when $j$ is the focal node. If $S\ne S'$, the same does not apply. Pair $\left(S,S'\right)$ is counted as such only when node $i$ is the focal node. When one of the neighbouring nodes $j$ is the focal node, then the same pair should be treated as $\left(S',S\right)$. A direct consequence is that
\begin{equation}
\left[S:S'\right]=\left[S':S\right],
\label{eqn:sim}
\end{equation}
just as formally stated above.

\paragraph*{\textbf{Global and local densities.}} With Eqs.~(\ref{eqn:konc1}-\ref{eqn:konc3}) in place, we can define global densities:
\begin{equation}
p_S=\frac{\left[S\right]}{N},\quad p_{SS'}=\frac{\left[S:S'\right]}{nN},
\end{equation}
and local densities:
\begin{equation}
q_{S|S'}=\frac{p_{SS'}}{p_{S'}},\quad q_{S|S'S''}=\frac{p_{SS'S''}}{p_{S'S''}},
\end{equation}
where $p_{S}$ represents the global proportion of nodes in state $S$, $p_{SS'}$ the global proportion of $SS'=\left(S,S'\right)$ pairs, $q_{S|S'}$ the average proportion of $S$-neighbours of  nodes in state $S'$, and $q_{S|S'S"}$ the average proportion of $S$-neighbours of $S'S''$ pairs.


Pair approximation is based on three conditions, (i) compatibility, (ii) symmetry, and (iii) closure \citep{szabo_pr07}. Compatibility with the mean-field theory is secured via
\begin{equation}
p_S=\sum_{S'}p_{SS'}
\end{equation}
which follows from Eq.~(\ref{eqn:konc5}), whereas symmetry
\begin{equation}
p_{SS'}=p_{S'S} \label{eqn:sim2}
\end{equation}
is a direct consequence of Eq.~(\ref{eqn:sim}). The concept of closure is somewhat more elaborate.

\begin{figure}[t!]
\centering
\includegraphics[scale=1.0]{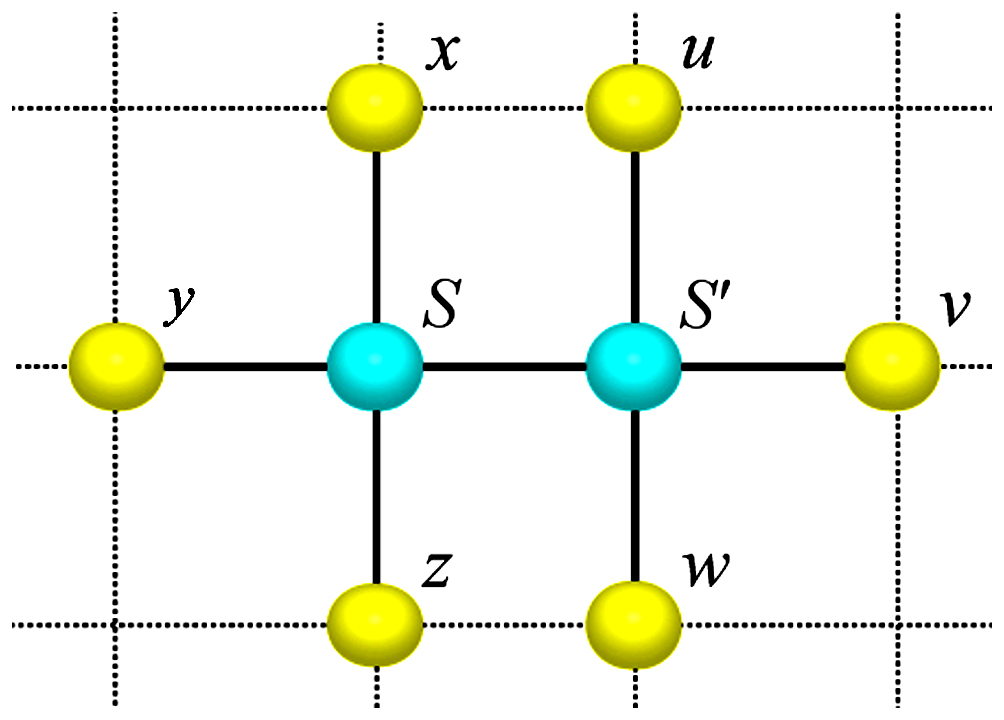}
\caption{Small part of a square lattice indicating the relevant configuration employed in the pair approximation technique. This configuration is used in determining pair transition probabilities $p_{SS'{\rightarrow}S'S'}$.}
\label{fig:8nodes}
\end{figure}

Achieving closure requires devising an approximation for proportions of larger node clusters. The ordinary pair approximation assumes that neighbours of neighbours have only minuscule effects no matter which focal node is looked at. This implies that
\begin{equation}
q_{S|S'S"} = q_{S|S'}. \label{eqn:clos1}
\end{equation}
Thus the probability of finding a triplet of nodes in state $SS'S''=\left(S,S',S''\right)$ is approximated as
\begin{align}
p_{SS'S''} &= q_{S|S'S''}\,p_{S'S''}= q_{S|S'}\,p_{S'S''}=  \nonumber\\
             &= \frac{p_{SS'}p_{S'S"}}{p_{S'}}. \label{eqn:clos2}
\end{align}
Furthermore, the probability of finding a cluster of eight nodes configured as shown in Fig.~\ref{fig:8nodes} becomes
\begin{eqnarray}
p_{xyzSS'uvw}=\frac{p_{xS}p_{yS}p_{zS}p_{SS'}p_{uS'}p_{vS'}p_{wS'}}{p_S^3p_{S'}^3}. \label{eqn:konfiguracija}
\end{eqnarray}



\paragraph*{\textbf{Adoption and transition probabilities.}} In evolutionary games with pairwise comparisons, a randomly chosen focal node, $i$, updates its strategy $S$ by comparing own payoff $P_S$ with payoff $P_{S'}$ of a randomly chosen neighbour, $j$, who opted for strategy $S'$. Node $i$ adopts strategy $S'$ with probability
\begin{equation}
W_{S{\leftarrow}S'} = \frac{1}{1+\exp{\left(\frac{P_{S}-P_{S'}}{K}\right)}}. \label{eqn:Fermi}
\end{equation}
Payoffs $P_{S}$ and $P_{S'}$ are accumulated in interactions with neighbouring nodes. Parameter $K$ quantifies the irrationality of selection, i.e., the larger the value of $K$, the greater is the probability of adopting $S'$ in favour of $S$ even if $P_{S}>P_{S'}$. Quantity $W_{S{\leftarrow}S'}$ is called the probability of adoption.

When deriving equations for pair approximation, it is generally a good idea to keep in mind information on local configurations of strategies. Following the notation in Fig.~\ref{fig:8nodes}, strategies neighbouring $S$ are $x$, $y$, $z$, and $S'$, while strategies neighbouring $S'$ are $u$, $v$, $w$, and $S$. For the purpose of keeping track of what goes on in the neighbourhoods of $S$ and $S'$, we introduce extended labels for adoption probabilities as follows
\begin{equation}
W_{S\leftarrow S'} = W(P_{S}^{xyz} \leftarrow P_{S'}^{uvw}). \label{eqn:Fermi2}
\end{equation}
Now we can calculate the probability that the pair of strategies $SS'$ transitions into $S'S'$:
\begin{align}
p_{SS'{\rightarrow}S'S'} &= \sum_{x,y,z}\sum_{u,v,w}  W\left(P_{S}^{xyz} \leftarrow P_{S'}^{uvw}\right) \nonumber \\
& \cdot\frac{p_{xS}p_{yS}p_{zS}p_{SS'}p_{uS'}p_{vS'}p_{wS'}}{p_S^3p_{S'}^3}.  \label{eqn:prijelaz}
\end{align}
Here the summations go over all possible configurations (Fig.~\ref{fig:8nodes}) whose probabilities are given by Eq.~(\ref{eqn:konfiguracija}). Quantity $p_{SS'\rightarrow S'S'}$ is called the probability of transition.


\begin{table}[b!]
  \begin{center}
    \footnotesize
\caption{The change in the number of $SS'$ pairs due to transitions $TT'{\rightarrow}T'T'$. Numbers $n_{S}=n_{S}^{xyz}$ and/or $n_{S'}=n_{S'}^{xyz}$ respectively signify the abundance of strategies $S$ and/or $S'$ among $x$, $y$, and $z$. Pairs $DC$, $LC$, $HC$, $LD$, $HD$, and $HL$ are left out due to the symmetry in Eq.~(\ref{eqn:sim2}).}
\begin{tabular}{lcccc}
\hline \hline
 $SS'$	 &$CD\rightarrow DD$	&$DC\rightarrow CD$  &$CL\rightarrow LL$ &$LC\rightarrow CC$ \\
\hline
$CC$	  &$-2n_{C}$	&$2(n_{C}+1)$	&$-2n_{C}$	&$2(n_{C}+1)$				 \\
$DD$	  &$2(n_{D}+1)$	&$-2n_{D}$	&$0$	 	&$0$			 \\
$LL$      &$0$		&$0$		&$2(n_{L}+1)$	&$-2n_{L}$	 \\
$HH$      &$0$		&$0$		&$0$		&$0$			 \\
$CD$      &$n_{C}-n_{D}-1$&$n_{D}-n_{C}-1$&$-n_{D}$	&$n_{D}$	 \\
$CL$      &$-n_{L}$	&$n_{L}$	&$n_{C}-n_{L}-1$&$n_{L}-n_{C}-1$	 \\
$CH$      &$-n_{H}$	&$n_{H}$	&$-n_{H}$ 	&$n_{H}$	\\
$DL$      &$n_{L}$	&$-n_{L}$	&$n_{D}$	&$-n_{D}$	\\
$DH$      &$n_{H}$	&$-n_{H}$	&$0$		&$0$		 \\
$LH$      &$0$		&$0$	 	&$n_{H}$	&$-n_{H}$	\\
\hline \hline
 $SS'$	 &$CH\rightarrow HH$ &$HC\rightarrow CC$ &$DL\rightarrow LL$ &$LD\rightarrow DD$ \\
\hline
$CC$	  &$-2n_{C}$	&$2(n_{C}+1)$	&$0$		&$0$				 \\
$DD$	  &$0$		&$0$	 	&$-2n_{D}$	&$2(n_{D}+1)$	\\
$LL$      &$0$		&$0$ 		&$2(n_{L}+1)$	&$-2n_{L}$	\\
$HH$      &$2(n_{H}+1)$	&$-n_{H}$	&$0$		&$0$			 \\
$CD$      &$-n_{D}$	&$n_{D}$	&$-n_{C}$	&$n_{C}$	 \\
$CL$      &$-n_{L}$	&$n_{L}$	&$n_{C}$	&$-n_{C}$		 \\
$CH$      &$n_{C}-n_{H}-1$&$n_{H}-n_{C}-1$ &$0$		&$0$		\\
$DL$      &$0$		&$0$		&$n_{D}-n_{L}-1$&$n_{L}-n_{D}-1$ \\
$DH$      &$n_{D}$	&$-n_{D}$	&$-n_{H}$	&$n_{H}$	 \\
$LH$      &$n_{L}$	&$-n_{L}$	&$n_{H}$	&$-n_{H}$	\\
\hline \hline
 $SS'$	  &$DH\rightarrow HH$ &$HD\rightarrow DD$ &$LH\rightarrow HH$ &$HL\rightarrow LL$ \\
\hline
$CC$	  &$0$		&$0$		&$0$		&$0$		 \\
$DD$	  &$-2n_{D}$	&$2(n_{D}+1)$	&$0$		&$0$ \\
$LL$      &$0$		&$0$		&$-2n_{L}$	&$2(n_{L}+1)$ \\
$HH$      &$2(n_{H}+1)$	&$-2n_{H}$	&$2(n_{H}+1)$	&$-2n_{H}$ \\
$CD$      &$-n_{C}$	&$n_{C}$	&$0$		&$0$		\\
$CL$      &$0$		&$0$		&$-n_{C}$	&$n_{C}$ \\
$CH$      &$n_{C}$	&$-n_{C}$	&$n_{C}$	&$-n_{C}$ \\
$DL$      &$-n_{L}$	&$n_{L}$	&$-n_{D}$	&$n_{D}$\\
$DH$      &$n_{D}-n_{H}-1$&$n_{H}-n_{D}-1$&$n_{D}$	&$-n_{D}$ \\
$LH$      &$n_{L}$	&$-n_{L}$	&$n_{L}-n_{H}-1$&$n_{H}-n_{L}-1$\\
\hline \hline
    \end{tabular}
\label{tab:koef}
  \end{center}
\end{table}

\paragraph*{\textbf{Pair equations.}} Let us for the moment focus on one specific pair of nodes whose strategies are $TT'$. When the node with strategy $T$ adopts the neighbour's strategy $T'$, $T{\ne}T'$, all possible global densities $p_{SS'}$ for which either $S$ or $S'\in\{T,T'\}$ undergo a change. To find how exactly global densities $p_{SS'}$ change in response to transition $TT'{\rightarrow}T'T'$, we need several ingredients.

The first ingredient is transition probability $p_{TT'{\rightarrow}T'T'}$. For any combination of strategies $T,T'{\in}\Sigma$, $T{\ne}T'$, the recipe for calculating transition probability $p_{TT'{\rightarrow}T'T'}$ is given by Eq.~(\ref{eqn:prijelaz}). If, for example, $T=C$ and $T'=D$, the transition probability from $CD$ to $DD$ is
\begin{align}
p_{CD{\rightarrow}DD} =& \sum_{x,y,z}\sum_{u,v,w}  W(P_{C}^{xyz} \leftarrow P_{D}^{uvw}) \nonumber \\
                       & \cdot\, \frac{p_{xC}p_{yC}p_{zC}p_{CD}p_{uD}p_{vD}p_{wD}}{p_C^3p_{D}^3}, \label{eqn:prijelazCD}
\end{align}
where $x$, $y$, $z$, and $D$, are the strategies neighbouring $C$, while $u$, $v$, $w$, and $C$ are the strategies neighbouring $D$. The probabilities for the remaining 11 transitions $TT'{\rightarrow}T'T'$ are obtainable in an entirely analogous manner.

Another ingredient in calculating how global densities $p_{SS'}$ change is to recognise that, in response to transition $TT'{\rightarrow}T'T'$, the local number of $SS'$ pairs, denoted ${\Delta}n_{SS'}$ changes as well. We express this latter change in terms of the abundance of strategy $S$ and/or $S'$ among $x$, $y$, and $z$, denoted respectively $n_S=n_S^{xyz}$ and $n_{S'}=n_{S'}^{xyz}$. Using the example from above, i.e., the transition from $CD$ to $DD$, we have ${\Delta}n_{CC}=-2n_C$. Table~\ref{tab:koef} lists all possible ${\Delta}n_{SS'}$ values in response to all 12 transitions of the form $TT'{\rightarrow}T'T'$.

We now have all the ingredients necessary to specify the dynamics of any global density $p_{SS'}$ brought about by transitions involving the $TT'$ pair. To do so, we need to calculate a weighted sum of changes in the number of $SS'$ pairs caused by transitions $TT'{\rightarrow}T'T'$, where the weights are the probabilities of each particular transition taking place. For example:
\[\begin{array}{l}
{\dot p}_{CC} = \\
\frac{2\,p_{CD}}{p_C^3p_D^3}\cdot
\left\{ \begin{array}{l}
	\sum\limits_{x,y,z} \left[n_C^{xyz} + 1\right]p_{Dx}p_{Dy}p_{Dz}\sum\limits_{u,v,w} p_{Cu}p_{Cv}p_{Cw} \\
	\cdot W(P_D^{xyz}\leftarrow P_C^{uvw}) \\ \\
      	- \sum\limits_{x,y,z} n_C^{xyz}p_{Cx}p_{Cy}p_{Cz} \sum\limits_{u,v,w} p_{Du}p_{Dv}p_{Dw} \\
        \cdot W(P_C^{xyz}\leftarrow P_D^{uvw})
\end{array} \right. \\ \\
+ \frac{2\,p_{CL}}{p_C^3p_L^3}\cdot
\left\{ \begin{array}{l}
	\sum\limits_{x,y,z} \left[n_C^{xyz} + 1\right]p_{Lx}p_{Ly}p_{Lz}\sum\limits_{u,v,w} p_{Cu}p_{Cv}p_{Cw} \\
	\cdot W(P_L^{xyz}\leftarrow P_C^{uvw}) \\ \\
      	- \sum\limits_{x,y,z} n_C^{xyz}p_{Cx}p_{Cy}p_{Cz} \sum\limits_{u,v,w} p_{Lu}p_{Lv}p_{Lw} \\
        \cdot W(P_C^{xyz}\leftarrow P_L^{uvw})
\end{array} \right. \\\\
+ \frac{2\,p_{CH}}{p_C^3p_H^3}\cdot
\left\{ \begin{array}{l}
	\sum\limits_{x,y,z} \left[n_C^{xyz} + 1\right]p_{Hx}p_{Hy}p_{Hz}\sum\limits_{u,v,w} p_{Cu}p_{Cv}p_{Cw} \\
	\cdot W(P_H^{xyz}\leftarrow P_C^{uvw}) \\ \\
      	- \sum\limits_{x,y,z} n_C^{xyz}p_{Cx}p_{Cy}p_{Cz} \sum\limits_{u,v,w} p_{Hu}p_{Hv}p_{Hw} \\
       	\cdot W(P_C^{xyz}\leftarrow P_H^{uvw})
\end{array} \right..\\\\
\end{array}\]

The dynamics of other pairs is obtainable in an analogous manner using Eq.~(\ref{eqn:prijelaz}) and Table~\ref{tab:koef}. Although in an evolutionary game with four strategies there are 16 pairs in total, we need to follow only ten because of the symmetry expressed by Eq.~(\ref{eqn:sim2}). One more pair could be eliminated by constraint
\begin{equation}
\sum_{S,S'\in \{C,D,L,H\}}p_{SS'} = 1
\end{equation}
if desired. We however keep track of the following 10 pairs:  $CC$, $DD$, $LL$, $HH$, $CD$, $CL$, $CH$, $DL$, $DH$, and $LH$. A full specification of the pair dynamics is given below the references.

%

\clearpage
\onecolumngrid

\phantom{Can't see this!}
\vfill

\begin{align}
{\dot p}_{DD} &=
\frac{2\,p_{CD}}{p_C^3p_D^3}\cdot
\left\{ \begin{array}{l}
	\sum\limits_{x,y,z} \left[n_D^{xyz} + 1\right]p_{Cx}p_{Cy}p_{Cz}\sum\limits_{u,v,w} p_{Du}p_{Dv}p_{Dw} \cdot W(P_C^{xyz}\leftarrow P_D^{uvw}) \\
      	- \sum\limits_{x,y,z} n_D^{xyz}p_{Dx}p_{Dy}p_{Dz} \sum\limits_{u,v,w} p_{Cu}p_{Cv}p_{Cw} \cdot W(P_D^{xyz}\leftarrow P_C^{uvw})
\end{array} \right. \nonumber \\
&+ \frac{2\,p_{DL}}{p_D^3p_L^3}\cdot
\left\{ \begin{array}{l}
	\sum\limits_{x,y,z} \left[n_D^{xyz} + 1\right]p_{Lx}p_{Ly}p_{Lz}\sum\limits_{u,v,w} p_{Du}p_{Dv}p_{Dw} \cdot W(P_L^{xyz}\leftarrow P_D^{uvw}) \\
      	- \sum\limits_{x,y,z} n_D^{xyz}p_{Dx}p_{Dy}p_{Dz} \sum\limits_{u,v,w} p_{Lu}p_{Lv}p_{Lw} \cdot W(P_D^{xyz}\leftarrow P_L^{uvw})
\end{array} \right. \nonumber \\
&+ \frac{2\,p_{DH}}{p_D^3p_H^3}\cdot
\left\{ \begin{array}{l}
	\sum\limits_{x,y,z} \left[n_D^{xyz} + 1\right]p_{Hx}p_{Hy}p_{Hz}\sum\limits_{u,v,w} p_{Du}p_{Dv}p_{Dw} \cdot W(P_H^{xyz}\leftarrow P_D^{uvw}) \\
      	- \sum\limits_{x,y,z} n_D^{xyz}p_{Dx}p_{Dy}p_{Dz} \sum\limits_{u,v,w} p_{Hu}p_{Hv}p_{Hw} \cdot W(P_D^{xyz}\leftarrow P_H^{uvw})
\end{array} \right. \nonumber
\end{align}

\vfill

\begin{align}
{\dot p}_{LL} &=
\frac{2\,p_{CL}}{p_C^3p_L^3}\cdot
\left\{ \begin{array}{l}
	\sum\limits_{x,y,z} \left[n_L^{xyz} + 1\right]p_{Cx}p_{Cy}p_{Cz}\sum\limits_{u,v,w} p_{Lu}p_{Lv}p_{Lw} \cdot W(P_C^{xyz}\leftarrow P_L^{uvw}) \\
      	- \sum\limits_{x,y,z} n_L^{xyz}p_{Lx}p_{Ly}p_{Lz} \sum\limits_{u,v,w} p_{Cu}p_{Cv}p_{Cw} \cdot W(P_L^{xyz}\leftarrow P_C^{uvw})
\end{array} \right. \nonumber \\
&+ \frac{2\,p_{DL}}{p_D^3p_L^3}\cdot
\left\{ \begin{array}{l}
	\sum\limits_{x,y,z} \left[n_L^{xyz} + 1\right]p_{Dx}p_{Dy}p_{Dz}\sum\limits_{u,v,w} p_{Lu}p_{Lv}p_{Lw} \cdot W(P_D^{xyz}\leftarrow P_L^{uvw}) \\
      	- \sum\limits_{x,y,z} n_L^{xyz}p_{Lx}p_{Ly}p_{Lz} \sum\limits_{u,v,w} p_{Du}p_{Dv}p_{Dw} \cdot W(P_L^{xyz}\leftarrow P_D^{uvw})
\end{array} \right. \nonumber \\
&+ \frac{2\,p_{LH}}{p_L^3p_H^3}\cdot
\left\{ \begin{array}{l}
	\sum\limits_{x,y,z} \left[n_L^{xyz} + 1\right]p_{Hx}p_{Hy}p_{Hz}\sum\limits_{u,v,w} p_{Lu}p_{Lv}p_{Lw} \cdot W(P_H^{xyz}\leftarrow P_L^{uvw}) \\
      	- \sum\limits_{x,y,z} n_L^{xyz}p_{Lx}p_{Ly}p_{Lz} \sum\limits_{u,v,w} p_{Hu}p_{Hv}p_{Hw} \cdot W(P_L^{xyz}\leftarrow P_H^{uvw})
\end{array} \right. \nonumber
\end{align}

\vfill

\begin{align}
{\dot p}_{HH} &=
\frac{2\,p_{CH}}{p_C^3p_H^3}\cdot
\left\{ \begin{array}{l}
	\sum\limits_{x,y,z} \left[n_H^{xyz} + 1\right]p_{Cx}p_{Cy}p_{Cz}\sum\limits_{u,v,w} p_{Hu}p_{Hv}p_{Hw} \cdot W(P_C^{xyz}\leftarrow P_H^{uvw}) \\
      	- \sum\limits_{x,y,z} n_H^{xyz}p_{Hx}p_{Hy}p_{Hz} \sum\limits_{u,v,w} p_{Cu}p_{Cv}p_{Cw} \cdot W(P_H^{xyz}\leftarrow P_C^{uvw})
\end{array} \right. \nonumber \\
&+ \frac{2\,p_{DH}}{p_D^3p_H^3}\cdot
\left\{ \begin{array}{l}
	\sum\limits_{x,y,z} \left[n_H^{xyz} + 1\right]p_{Dx}p_{Dy}p_{Dz}\sum\limits_{u,v,w} p_{Hu}p_{Hv}p_{Hw} \cdot W(P_D^{xyz}\leftarrow P_H^{uvw}) \\
      	- \sum\limits_{x,y,z} n_H^{xyz}p_{Hx}p_{Hy}p_{Hz} \sum\limits_{u,v,w} p_{Du}p_{Dv}p_{Dw} \cdot W(P_H^{xyz}\leftarrow P_D^{uvw})
\end{array} \right. \nonumber \\
&+ \frac{2\,p_{LH}}{p_L^3p_H^3}\cdot
\left\{ \begin{array}{l}
	\sum\limits_{x,y,z} \left[n_H^{xyz} + 1\right]p_{Lx}p_{Ly}p_{Lz}\sum\limits_{u,v,w} p_{Hu}p_{Hv}p_{Hw} \cdot W(P_L^{xyz}\leftarrow P_H^{uvw}) \\
      	- \sum\limits_{x,y,z} n_H^{xyz}p_{Hx}p_{Hy}p_{Hz} \sum\limits_{u,v,w} p_{Lu}p_{Lv}p_{Lw} \cdot W(P_H^{xyz}\leftarrow P_L^{uvw})
\end{array} \right. \nonumber
\end{align}

\vfill

\clearpage

\phantom{Can't see this!}
\vfill

\begin{align}
{\dot p}_{CD} &=
\frac{2\,p_{CD}}{p_C^3p_D^3}\cdot
\left\{ \begin{array}{l}
	\sum\limits_{x,y,z} \frac{n_C^{xyz} - n_D^{xyz} - 1}{2}p_{Cx}p_{Cy}p_{Cz}\sum\limits_{u,v,w} p_{Du}p_{Dv}p_{Dw} \cdot W(P_C^{xyz}\leftarrow P_D^{uvw}) \\
      	+ \sum\limits_{x,y,z} \frac{n_D^{xyz} - n_C^{xyz} - 1}{2}p_{Dx}p_{Dy}p_{Dz} \sum\limits_{u,v,w} p_{Cu}p_{Cv}p_{Cw} \cdot W(P_D^{xyz}\leftarrow P_C^{uvw})
\end{array} \right. \nonumber \\
&+ \frac{2\,p_{CL}}{p_C^3p_L^3}\cdot
\left\{ \begin{array}{l}
	\sum\limits_{x,y,z} \frac{n_D^{xyz}}{2} p_{Lx}p_{Ly}p_{Lz}\sum\limits_{u,v,w} p_{Cu}p_{Cv}p_{Cw} \cdot W(P_L^{xyz}\leftarrow P_C^{uvw}) \\
      	- \sum\limits_{x,y,z} \frac{n_D^{xyz}}{2} p_{Cx}p_{Cy}p_{Cz} \sum\limits_{u,v,w} p_{Lu}p_{Lv}p_{Lw} \cdot W(P_C^{xyz}\leftarrow P_L^{uvw})
\end{array} \right. \nonumber \\
&+ \frac{2\,p_{CH}}{p_C^3p_H^3}\cdot
\left\{ \begin{array}{l}
	\sum\limits_{x,y,z} \frac{n_D^{xyz}}{2} p_{Hx}p_{Hy}p_{Hz}\sum\limits_{u,v,w} p_{Cu}p_{Cv}p_{Cw} \cdot W(P_H^{xyz}\leftarrow P_C^{uvw}) \\
      	- \sum\limits_{x,y,z} \frac{n_D^{xyz}}{2} p_{Cx}p_{Cy}p_{Cz} \sum\limits_{u,v,w} p_{Hu}p_{Hv}p_{Hw} \cdot W(P_C^{xyz}\leftarrow P_H^{uvw})
\end{array} \right. \nonumber \\
&+ \frac{2\,p_{DL}}{p_D^3p_L^3}\cdot
\left\{ \begin{array}{l}
	\sum\limits_{x,y,z} \frac{n_C^{xyz}}{2} p_{Lx}p_{Ly}p_{Lz}\sum\limits_{u,v,w} p_{Du}p_{Dv}p_{Dw} \cdot W(P_L^{xyz}\leftarrow P_D^{uvw}) \\
      	- \sum\limits_{x,y,z} \frac{n_C^{xyz}}{2} p_{Dx}p_{Dy}p_{Dz} \sum\limits_{u,v,w} p_{Lu}p_{Lv}p_{Lw} \cdot W(P_D^{xyz}\leftarrow P_L^{uvw})
\end{array} \right. \nonumber \\
&+ \frac{2\,p_{DH}}{p_D^3p_H^3}\cdot
\left\{ \begin{array}{l}
	\sum\limits_{x,y,z} \frac{n_C^{xyz}}{2}  p_{Hx}p_{Hy}p_{Hz}\sum\limits_{u,v,w} p_{Du}p_{Dv}p_{Dw} \cdot W(P_H^{xyz}\leftarrow P_D^{uvw}) \\
      	- \sum\limits_{x,y,z} \frac{n_C^{xyz}}{2} p_{Dx}p_{Dy}p_{Dz} \sum\limits_{u,v,w} p_{Hu}p_{Hv}p_{Hw} \cdot W(P_D^{xyz}\leftarrow P_H^{uvw})
\end{array} \right. \nonumber
\end{align}

\vfill

\begin{align}
{\dot p}_{CL} &=
\frac{2\,p_{CL}}{p_C^3p_L^3}\cdot
\left\{ \begin{array}{l}
	\sum\limits_{x,y,z} \frac{n_C^{xyz} - n_L^{xyz} - 1}{2}p_{Cx}p_{Cy}p_{Cz}\sum\limits_{u,v,w} p_{Lu}p_{Lv}p_{Lw} \cdot W(P_C^{xyz}\leftarrow P_L^{uvw}) \\
      	+ \sum\limits_{x,y,z} \frac{n_L^{xyz} - n_C^{xyz} - 1}{2}p_{Lx}p_{Ly}p_{Lz} \sum\limits_{u,v,w} p_{Cu}p_{Cv}p_{Cw} \cdot W(P_L^{xyz}\leftarrow P_C^{uvw})
\end{array} \right. \nonumber \\
&+ \frac{2\,p_{CD}}{p_C^3p_D^3}\cdot
\left\{ \begin{array}{l}
	\sum\limits_{x,y,z} \frac{n_L^{xyz}}{2} p_{Dx}p_{Dy}p_{Dz}\sum\limits_{u,v,w} p_{Cu}p_{Cv}p_{Cw} \cdot W(P_D^{xyz}\leftarrow P_C^{uvw}) \\
      	- \sum\limits_{x,y,z} \frac{n_L^{xyz}}{2} p_{Cx}p_{Cy}p_{Cz} \sum\limits_{u,v,w} p_{Du}p_{Dv}p_{Dw} \cdot W(P_C^{xyz}\leftarrow P_D^{uvw})
\end{array} \right. \nonumber \\
&+ \frac{2\,p_{CH}}{p_C^3p_H^3}\cdot
\left\{ \begin{array}{l}
	\sum\limits_{x,y,z} \frac{n_L^{xyz}}{2} p_{Hx}p_{Hy}p_{Hz}\sum\limits_{u,v,w} p_{Cu}p_{Cv}p_{Cw} \cdot W(P_H^{xyz}\leftarrow P_C^{uvw}) \\
      	- \sum\limits_{x,y,z} \frac{n_L^{xyz}}{2} p_{Cx}p_{Cy}p_{Cz} \sum\limits_{u,v,w} p_{Hu}p_{Hv}p_{Hw} \cdot W(P_C^{xyz}\leftarrow P_H^{uvw})
\end{array} \right. \nonumber \\
& + \frac{2\,p_{DL}}{p_D^3p_L^3}\cdot
\left\{ \begin{array}{l}
      	\sum\limits_{x,y,z} \frac{n_C^{xyz}}{2} p_{Dx}p_{Dy}p_{Dz} \sum\limits_{u,v,w} p_{Lu}p_{Lv}p_{Lw} \cdot W(P_D^{xyz}\leftarrow P_L^{uvw}) \\
	- \sum\limits_{x,y,z} \frac{n_C^{xyz}}{2} p_{Lx}p_{Ly}p_{Lz}\sum\limits_{u,v,w} p_{Du}p_{Dv}p_{Dw} \cdot W(P_L^{xyz}\leftarrow P_D^{uvw})
\end{array} \right. \nonumber \\
&+ \frac{2\,p_{LH}}{p_L^3p_H^3}\cdot
\left\{ \begin{array}{l}
	\sum\limits_{x,y,z} \frac{n_C^{xyz}}{2}  p_{Hx}p_{Hy}p_{Hz}\sum\limits_{u,v,w} p_{Lu}p_{Lv}p_{Lw} \cdot W(P_H^{xyz}\leftarrow P_L^{uvw}) \\
      	- \sum\limits_{x,y,z} \frac{n_C^{xyz}}{2} p_{Lx}p_{Ly}p_{Lz} \sum\limits_{u,v,w} p_{Hu}p_{Hv}p_{Hw} \cdot W(P_L^{xyz}\leftarrow P_H^{uvw})
\end{array} \right. \nonumber
\end{align}

\vfill

\clearpage

\phantom{Can't see this!}
\vfill

\begin{align}
{\dot p}_{CH} &=
\frac{2\,p_{CH}}{p_C^3p_H^3}\cdot
\left\{ \begin{array}{l}
	\sum\limits_{x,y,z} \frac{n_C^{xyz} - n_H^{xyz} - 1}{2}p_{Cx}p_{Cy}p_{Cz}\sum\limits_{u,v,w} p_{Hu}p_{Hv}p_{Hw} \cdot W(P_C^{xyz}\leftarrow P_H^{uvw}) \\
      	+ \sum\limits_{x,y,z} \frac{n_H^{xyz} - n_C^{xyz} - 1}{2}p_{Hx}p_{Hy}p_{Hz} \sum\limits_{u,v,w} p_{Cu}p_{Cv}p_{Cw} \cdot W(P_L^{xyz}\leftarrow P_C^{uvw})
\end{array} \right. \nonumber \\
&+ \frac{2\,p_{CD}}{p_C^3p_D^3}\cdot
\left\{ \begin{array}{l}
	\sum\limits_{x,y,z} \frac{n_H^{xyz}}{2} p_{Dx}p_{Dy}p_{Dz}\sum\limits_{u,v,w} p_{Cu}p_{Cv}p_{Cw} \cdot W(P_D^{xyz}\leftarrow P_C^{uvw}) \\
      	- \sum\limits_{x,y,z} \frac{n_H^{xyz}}{2} p_{Cx}p_{Cy}p_{Cz} \sum\limits_{u,v,w} p_{Du}p_{Dv}p_{Dw} \cdot W(P_C^{xyz}\leftarrow P_D^{uvw})
\end{array} \right. \nonumber \\
&+ \frac{2\,p_{CL}}{p_C^3p_L^3}\cdot
\left\{ \begin{array}{l}
	\sum\limits_{x,y,z} \frac{n_H^{xyz}}{2} p_{Lx}p_{Ly}p_{Lz}\sum\limits_{u,v,w} p_{Cu}p_{Cv}p_{Cw} \cdot W(P_L^{xyz}\leftarrow P_C^{uvw}) \\
      	- \sum\limits_{x,y,z} \frac{n_H^{xyz}}{2} p_{Cx}p_{Cy}p_{Cz} \sum\limits_{u,v,w} p_{Lu}p_{Lv}p_{Lw} \cdot W(P_C^{xyz}\leftarrow P_L^{uvw})
\end{array} \right. \nonumber \\
& + \frac{2\,p_{DH}}{p_D^3p_H^3}\cdot
\left\{ \begin{array}{l}
      	\sum\limits_{x,y,z} \frac{n_C^{xyz}}{2} p_{Dx}p_{Dy}p_{Dz} \sum\limits_{u,v,w} p_{Hu}p_{Hv}p_{Hw} \cdot W(P_D^{xyz}\leftarrow P_H^{uvw}) \\
	- \sum\limits_{x,y,z} \frac{n_C^{xyz}}{2} p_{Hx}p_{Hy}p_{Hz}\sum\limits_{u,v,w} p_{Du}p_{Dv}p_{Dw} \cdot W(P_H^{xyz}\leftarrow P_D^{uvw})
\end{array} \right. \nonumber \\
&+ \frac{2\,p_{LH}}{p_L^3p_H^3}\cdot
\left\{ \begin{array}{l}
	\sum\limits_{x,y,z} \frac{n_C^{xyz}}{2} p_{Lx}p_{Ly}p_{Lz} \sum\limits_{u,v,w} p_{Hu}p_{Hv}p_{Hw} \cdot W(P_L^{xyz}\leftarrow P_H^{uvw}) \\
      	- \sum\limits_{x,y,z} \frac{n_C^{xyz}}{2}  p_{Hx}p_{Hy}p_{Hz}\sum\limits_{u,v,w} p_{Lu}p_{Lv}p_{Lw} \cdot W(P_H^{xyz}\leftarrow P_L^{uvw})
\end{array} \right. \nonumber
\end{align}

\vfill

\begin{align}
{\dot p}_{DL} &=
\frac{2\,p_{DL}}{p_D^3p_L^3}\cdot
\left\{ \begin{array}{l}
	\sum\limits_{x,y,z} \frac{n_D^{xyz} - n_L^{xyz} - 1}{2}p_{Dx}p_{Dy}p_{Dz}\sum\limits_{u,v,w} p_{Lu}p_{Lv}p_{Lw} \cdot W(P_D^{xyz}\leftarrow P_L^{uvw}) \\
      	+ \sum\limits_{x,y,z} \frac{n_L^{xyz} - n_D^{xyz} - 1}{2}p_{Lx}p_{Ly}p_{Lz} \sum\limits_{u,v,w} p_{Du}p_{Dv}p_{Dw} \cdot W(P_L^{xyz}\leftarrow P_D^{uvw})
\end{array} \right. \nonumber \\
&+ \frac{2\,p_{CD}}{p_C^3p_D^3}\cdot
\left\{ \begin{array}{l}
	\sum\limits_{x,y,z} \frac{n_L^{xyz}}{2} p_{Cx}p_{Cy}p_{Cz} \sum\limits_{u,v,w} p_{Du}p_{Dv}p_{Dw} \cdot W(P_C^{xyz}\leftarrow P_D^{uvw}) \\
      	- \sum\limits_{x,y,z} \frac{n_L^{xyz}}{2} p_{Dx}p_{Dy}p_{Dz}\sum\limits_{u,v,w} p_{Cu}p_{Cv}p_{Cw} \cdot W(P_D^{xyz}\leftarrow P_C^{uvw})
\end{array} \right. \nonumber \\
&+ \frac{2\,p_{CL}}{p_C^3p_L^3}\cdot
\left\{ \begin{array}{l}
	\sum\limits_{x,y,z} \frac{n_D^{xyz}}{2} p_{Cx}p_{Cy}p_{Cz} \sum\limits_{u,v,w} p_{Lu}p_{Lv}p_{Lw} \cdot W(P_C^{xyz}\leftarrow P_L^{uvw}) \\
      	- \sum\limits_{x,y,z} \frac{n_D^{xyz}}{2} p_{Lx}p_{Ly}p_{Lz}\sum\limits_{u,v,w} p_{Cu}p_{Cv}p_{Cw} \cdot W(P_L^{xyz}\leftarrow P_C^{uvw})
\end{array} \right. \nonumber \\
& + \frac{2\,p_{DH}}{p_D^3p_H^3}\cdot
\left\{ \begin{array}{l}
      	\sum\limits_{x,y,z} \frac{n_L^{xyz}}{2} p_{Hx}p_{Hy}p_{Hz}\sum\limits_{u,v,w} p_{Du}p_{Dv}p_{Dw} \cdot W(P_H^{xyz}\leftarrow P_D^{uvw}) \\
	- \sum\limits_{x,y,z} \frac{n_L^{xyz}}{2} p_{Dx}p_{Dy}p_{Dz} \sum\limits_{u,v,w} p_{Hu}p_{Hv}p_{Hw} \cdot W(P_D^{xyz}\leftarrow P_H^{uvw})
\end{array} \right. \nonumber \\
&+ \frac{2\,p_{LH}}{p_L^3p_H^3}\cdot
\left\{ \begin{array}{l}
	\sum\limits_{x,y,z} \frac{n_D^{xyz}}{2}  p_{Hx}p_{Hy}p_{Hz}\sum\limits_{u,v,w} p_{Lu}p_{Lv}p_{Lw} \cdot W(P_H^{xyz}\leftarrow P_L^{uvw}) \\
      	- \sum\limits_{x,y,z} \frac{n_D^{xyz}}{2} p_{Lx}p_{Ly}p_{Lz} \sum\limits_{u,v,w} p_{Hu}p_{Hv}p_{Hw} \cdot W(P_L^{xyz}\leftarrow P_H^{uvw})
\end{array} \right. \nonumber
\end{align}

\vfill

\clearpage

\phantom{Can't see this!}
\vfill

\begin{align}
{\dot p}_{DH} &=
\frac{2\,p_{DH}}{p_D^3p_H^3}\cdot
\left\{ \begin{array}{l}
	\sum\limits_{x,y,z} \frac{n_D^{xyz} - n_H^{xyz} - 1}{2}p_{Dx}p_{Dy}p_{Dz}\sum\limits_{u,v,w} p_{Hu}p_{Hv}p_{Hw} \cdot W(P_D^{xyz}\leftarrow P_H^{uvw}) \\
      	+ \sum\limits_{x,y,z} \frac{n_H^{xyz} - n_D^{xyz} - 1}{2}p_{Hx}p_{Hy}p_{Hz} \sum\limits_{u,v,w} p_{Du}p_{Dv}p_{Dw} \cdot W(P_H^{xyz}\leftarrow P_D^{uvw})
\end{array} \right. \nonumber \\
&+ \frac{2\,p_{CD}}{p_C^3p_D^3}\cdot
\left\{ \begin{array}{l}
	\sum\limits_{x,y,z} \frac{n_H^{xyz}}{2} p_{Cx}p_{Cy}p_{Cz} \sum\limits_{u,v,w} p_{Du}p_{Dv}p_{Dw} \cdot W(P_C^{xyz}\leftarrow P_D^{uvw}) \\
      	- \sum\limits_{x,y,z} \frac{n_H^{xyz}}{2} p_{Dx}p_{Dy}p_{Dz}\sum\limits_{u,v,w} p_{Cu}p_{Cv}p_{Cw} \cdot W(P_D^{xyz}\leftarrow P_C^{uvw})
\end{array} \right. \nonumber \\
&+ \frac{2\,p_{CH}}{p_C^3p_H^3}\cdot
\left\{ \begin{array}{l}
	\sum\limits_{x,y,z} \frac{n_D^{xyz}}{2} p_{Cx}p_{Cy}p_{Cz} \sum\limits_{u,v,w} p_{Hu}p_{Hv}p_{Hw} \cdot W(P_C^{xyz}\leftarrow P_H^{uvw}) \\
      	- \sum\limits_{x,y,z} \frac{n_D^{xyz}}{2} p_{Hx}p_{Hy}p_{Hz}\sum\limits_{u,v,w} p_{Cu}p_{Cv}p_{Cw} \cdot W(P_H^{xyz}\leftarrow P_C^{uvw})
\end{array} \right. \nonumber \\
& + \frac{2\,p_{DL}}{p_D^3p_L^3}\cdot
\left\{ \begin{array}{l}
      	\sum\limits_{x,y,z} \frac{n_H^{xyz}}{2} p_{Lx}p_{Ly}p_{Lz}\sum\limits_{u,v,w} p_{Du}p_{Dv}p_{Dw} \cdot W(P_L^{xyz}\leftarrow P_D^{uvw}) \\
	- \sum\limits_{x,y,z} \frac{n_H^{xyz}}{2} p_{Dx}p_{Dy}p_{Dz} \sum\limits_{u,v,w} p_{Lu}p_{Lv}p_{Lw} \cdot W(P_D^{xyz}\leftarrow P_L^{uvw})
\end{array} \right. \nonumber \\
&+ \frac{2\,p_{LH}}{p_L^3p_H^3}\cdot
\left\{ \begin{array}{l}
	\sum\limits_{x,y,z} \frac{n_D^{xyz}}{2} p_{Lx}p_{Ly}p_{Lz} \sum\limits_{u,v,w} p_{Hu}p_{Hv}p_{Hw} \cdot W(P_L^{xyz}\leftarrow P_H^{uvw}) \\
      	- \sum\limits_{x,y,z} \frac{n_D^{xyz}}{2}  p_{Hx}p_{Hy}p_{Hz}\sum\limits_{u,v,w} p_{Lu}p_{Lv}p_{Lw} \cdot W(P_H^{xyz}\leftarrow P_L^{uvw})
\end{array} \right. \nonumber
\end{align}

\vfill

\begin{align}
{\dot p}_{LH} &=
\frac{2\,p_{LH}}{p_L^3p_H^3}\cdot
\left\{ \begin{array}{l}
	\sum\limits_{x,y,z} \frac{n_L^{xyz} - n_H^{xyz} - 1}{2}p_{Lx}p_{Ly}p_{Lz}\sum\limits_{u,v,w} p_{Hu}p_{Hv}p_{Hw} \cdot W(P_L^{xyz}\leftarrow P_H^{uvw}) \\
      	+ \sum\limits_{x,y,z} \frac{n_H^{xyz} - n_L^{xyz} - 1}{2}p_{Hx}p_{Hy}p_{Hz} \sum\limits_{u,v,w} p_{Lu}p_{Lv}p_{Lw} \cdot W(P_H^{xyz}\leftarrow P_L^{uvw})
\end{array} \right. \nonumber \\
&+ \frac{2\,p_{CL}}{p_C^3p_L^3}\cdot
\left\{ \begin{array}{l}
	\sum\limits_{x,y,z} \frac{n_H^{xyz}}{2} p_{Cx}p_{Cy}p_{Cz} \sum\limits_{u,v,w} p_{Lu}p_{Lv}p_{Lw} \cdot W(P_C^{xyz}\leftarrow P_L^{uvw}) \\
      	- \sum\limits_{x,y,z} \frac{n_H^{xyz}}{2} p_{Lx}p_{Ly}p_{Lz}\sum\limits_{u,v,w} p_{Cu}p_{Cv}p_{Cw} \cdot W(P_L^{xyz}\leftarrow P_C^{uvw})
\end{array} \right. \nonumber \\
&+ \frac{2\,p_{CH}}{p_C^3p_H^3}\cdot
\left\{ \begin{array}{l}
	\sum\limits_{x,y,z} \frac{n_L^{xyz}}{2} p_{Cx}p_{Cy}p_{Cz} \sum\limits_{u,v,w} p_{Hu}p_{Hv}p_{Hw} \cdot W(P_C^{xyz}\leftarrow P_H^{uvw}) \\
      	- \sum\limits_{x,y,z} \frac{n_L^{xyz}}{2} p_{Hx}p_{Hy}p_{Hz}\sum\limits_{u,v,w} p_{Cu}p_{Cv}p_{Cw} \cdot W(P_H^{xyz}\leftarrow P_C^{uvw})
\end{array} \right. \nonumber \\
& + \frac{2\,p_{DL}}{p_D^3p_L^3}\cdot
\left\{ \begin{array}{l}
      	\sum\limits_{x,y,z} \frac{n_H^{xyz}}{2} p_{Dx}p_{Dy}p_{Dz} \sum\limits_{u,v,w} p_{Lu}p_{Lv}p_{Lw} \cdot W(P_D^{xyz}\leftarrow P_L^{uvw}) \\
	- \sum\limits_{x,y,z} \frac{n_H^{xyz}}{2} p_{Lx}p_{Ly}p_{Lz}\sum\limits_{u,v,w} p_{Du}p_{Dv}p_{Dw} \cdot W(P_L^{xyz}\leftarrow P_D^{uvw})
\end{array} \right. \nonumber \\
&+ \frac{2\,p_{DH}}{p_D^3p_H^3}\cdot
\left\{ \begin{array}{l}
	\sum\limits_{x,y,z} \frac{n_L^{xyz}}{2} p_{Dx}p_{Dy}p_{Dz} \sum\limits_{u,v,w} p_{Hu}p_{Hv}p_{Hw} \cdot W(P_D^{xyz}\leftarrow P_H^{uvw}) \\
      	- \sum\limits_{x,y,z} \frac{n_L^{xyz}}{2}  p_{Hx}p_{Hy}p_{Hz}\sum\limits_{u,v,w} p_{Du}p_{Dv}p_{Dw} \cdot W(P_H^{xyz}\leftarrow P_D^{uvw})
\end{array} \right. \nonumber
\end{align}

\vfill

\end{document}